\documentclass[a4paper,prx, superscriptaddress, twocolumn,longbibliography]{revtex4-2}

\usepackage{microtype}

\usepackage{silence}
\usepackage[utf8]{inputenc}
\usepackage[english]{babel}
\usepackage[pdftex,dvipsnames]{xcolor}     
\usepackage{graphicx}
\usepackage{amsmath,amssymb}
\usepackage{amsfonts}
\usepackage{mathtools}
\usepackage{braket}
\renewcommand{\eqref}[1]{(\ref{#1})}

\usepackage{titlesec}
\titleformat*{\section}{\raggedright\bfseries\sffamily\large}
\titleformat*{\subsection}{\raggedright\bfseries\sffamily}
\titleformat*{\subsubsection}{\raggedright\bfseries\sffamily}

\makeatletter
\renewcommand{\fnum@figure}{\textbf{Fig.~\thefigure}}
\makeatother
\usepackage{caption}
\usepackage{bm}
\renewcommand{\vec}{\ensuremath{\bm}}

\renewcommand{\Im}{\text{Im}}
\renewcommand{\Re}{\text{Re}}
\definecolor{bleu}{RGB}{0,47,108}
\usepackage{hyperref}
\hypersetup{colorlinks,citecolor={blue},linkcolor={red},urlcolor={red}}

\renewcommand{\eqref}[1]{(\ref{#1})}
\usepackage{enumitem}
\usepackage{pgfornament}
\usepackage{empheq}

\usepackage{xargs} 
\usepackage[colorinlistoftodos,prependcaption,textsize=footnotesize]{todonotes}
\newcommandx{\unsure}[2][1=]{\todo[linecolor=red,backgroundcolor=red!25,bordercolor=red,#1]{#2}}
\newcommandx{\change}[2][1=]{\todo[linecolor=blue,backgroundcolor=blue!25,bordercolor=blue,#1]{#2}}
\newcommandx{\info}[2][1=]{\todo[linecolor=OliveGreen,backgroundcolor=OliveGreen!25,bordercolor=OliveGreen,#1]{#2}}
\newcommandx{\improvement}[2][1=]{\todo[linecolor=Plum,backgroundcolor=Plum!25,bordercolor=Plum,#1]{#2}}
\newcommandx{\thiswillnotshow}[2][1=]{\todo[disable,#1]{#2}}

\usepackage{nicematrix}

\begin{document}

\title{\sffamily\Large Morphologies of caustics studied by catastrophe charged-particle optics}
\date{\today}

\author{Tom Fraysse}
\affiliation{CEMES-CNRS, Université de Toulouse, CNRS, Toulouse, France}

\author{Robin Cours}
\affiliation{CEMES-CNRS, Université de Toulouse, CNRS, Toulouse, France}

\author{Hugo Lourenço-Martins}\email[]{hugo.lourenco-martins@cnrs.fr}
\affiliation{CEMES-CNRS, Université de Toulouse, CNRS, Toulouse, France}

\author{Florent Houdellier}\email[]{florent.houdellier@cnrs.fr}
\affiliation{CEMES-CNRS, Université de Toulouse, CNRS, Toulouse, France}

\begin{abstract}
This paper explores the topologies of caustics observed in instruments that employ charged particles, such as electron and ion microscopes. These geometrical figures are studied here using catastrophe theory. The application of this geometrical theory to our optical situation has enabled us to analytically reproduce the behaviours of various caustics. The interest lies mainly in the universal nature of these results since our treatment requires no prior knowledge of the optical configuration, but only a smart definition of the control space. This universal approach has finally made it possible to extract mathematical relationships between the aberration coefficients of any optical system, which were hidden by the complexity of optical trajectories but revealed by the set of catastrophes in the control space. These results provide a glimpse for future applications of caustics in the development of new corrected optical systems, especially for ions-based devices. 
\end{abstract}

\maketitle


\section{Introduction}
\begin{figure}
\centering
\includegraphics[width=0.5\textwidth]{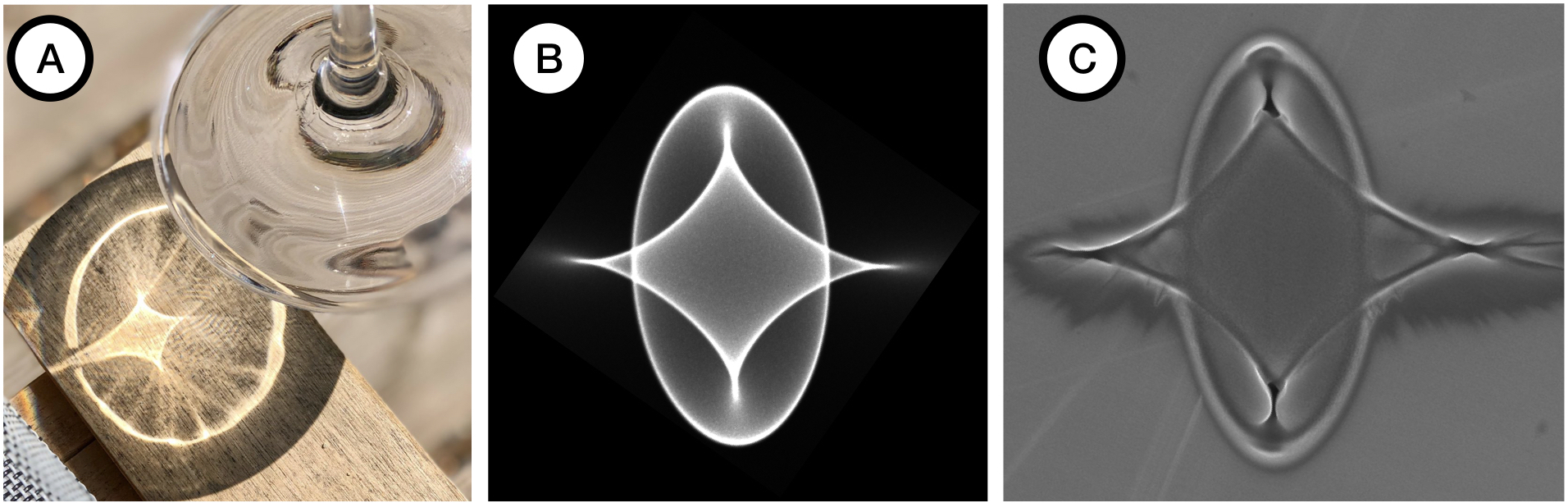}
\caption{\label{fig1} A caustic observed with (a) a light beam (b) electrons and (c) ions beams}
\end{figure}

Everyone has experienced the pleasure of seeing the formation of elaborate geometric shapes from light, whether on a table, a wall, or any other surface after a sunbeam has passed through a transparent material (see Fig.~\ref{fig1}(a)). These mesmerizing figures of diverse shapes - commonly referred to as \emph{caustics} - leave no one indifferent \cite{hesselgren_architectural_2013}. Their profound beauty is further revealed when working with charged-particle devices, such as electron or ion microscopes, where similar complex caustic formations can be observed. Indeed, the observation of the same optical phenomenon with particles of such a different nature - light, electron, or ion (see Fig.~\ref{fig1}(a-c)) - points towards the existence of a universal underlying principle.\\

The Hamiltonian picture of optics and mechanics constitutes an adapted framework for uncovering this principle. Within this framework, the behaviour of any optical or mechanical system can be derived from the mathematical properties of a single function called the \textit{characteristic function} through the principle of least action \cite{hamilton_general_1833, martin-robine_histoire_2006} -  the main challenge lying in determining the appropriate characteristic function.

In conventional (light) optics, an appropriate choice for this function corresponds to the optical path length $l$, as suggested by the Fermat's principle \cite{synge_geometrical_1937}. In that case, the trajectory followed by the light is given by the extremum of $l$:

\begin{equation}\label{fermatderivationcaustic}
\frac{\partial l}{\partial x}=\frac{\partial l}{\partial y}=0
\end{equation}

\noindent where the variables $x,y$ label the plane transverse to the ray trajectory taken along the optical axis $z$ \cite{buchdahl_introduction_1970,luneburg_mathematical_2021}. The caustics then correspond to the regions of space $(x,y)$ where a high concentration of trajectories that have various initial conditions converge, i.e. have the same function $l$ \cite{nye_natural_1999}. Thus, outside the caustic region $l$ is single valued (there is one ray through each point), while inside the caustic $l$ is $n$-valued ($n$ rays through each point). In other words, the degenerative set of the multi-valued characteristic function will correspond to the location of the caustic surface.\\

Hamiltonian optics therefore provides a direct operational definition of caustics through the application of the stationarity principle \eqref{fermatderivationcaustic}, regardless of the physical situation studied \cite{luneburg_mathematical_2021,cline_variational_2019}. Nonetheless, this approach does not explicit the underlying principle connecting the geometry of caustics to the structure of this multi-valued characteristic functions. In particular, it does not explain why different characteristic functions describing distinct physical systems - may they be mechanical or optical - would lead to the same caustic figures. This is where \emph{catastrophe theory} becomes relevant.\\
Developed by Rene Thom, this theory deals with the topological structure of smooth functions and their singularities \cite{thom_stabilite_1974,arnold_catastrophe_1992,zeeman_catastrophe_1977}. Catastrophe theory has encountered a great success thanks to its efficiency in predicting discontinuous behaviour in systems described by a characteristic function, which span a wide range of practical situations, e.g. the buckling of bridges, liquid-gas phase transition, spin transition in the Ising model or even cellular differentiation \cite{poston_catastrophe_1996}. More generally, it applies to almost any situation encountered in modern physics studied under the Hamiltonian pictures, hence to caustics formation \cite{berry_iv_1980, selmke_bubble_2020}.

The goal of this paper is to apply catastrophe theory to the study of caustics formation in charged particle optics (CPO) systems. In section \ref{SEC:introduction_catastrophe_theory}, we start by providing a minimal introduction to catastrophe theory and its application to light optics. In section \ref{sec:OPC}, through the application of the principle of least action, we will derive the characteristic function of CPO following the early work of \cite{rose_geometrical_2009}. This will illustrate the great advantage of the Hamiltonian approach, as it enables us to use a common language for charged particles and conventional optics \cite{torre_linear_2005}. The physical concepts classically derived from this language (paraxial optics, aberrations theory for instance) will then be identical in all these situations. In section \ref{catastropheCPO}, we will apply catastrophe theory to the characteristic function derived in section \ref{sec:OPC}. It will lead to the definition of a catastrophe charged-particle optics (CCPO) framework, in which the connection between the structural properties of the characteristic function and the geometry of the caustics become transparent. Eventually, in sections \ref{sec:application1} and \ref{sec:application2}, we will apply this formalism to derive quantitative relations between the aberration coefficients and the geometrical properties of simple caustics. This final section will highlight the strong potential of CCPO for fast and efficient measurement of the aberration coefficients of CPO systems.

\section{A short introduction to catastrophe theory} \label{SEC:introduction_catastrophe_theory}

To be able to mathematically study any physical system using the catastrophe theory, we have to define a so-called generating function $\Phi (X,C)$ \cite{barbero_catastrophe_2022}, which is linked - but not necessarily equal - to the characteristic function. In the following, we will separate the variables of $\Phi$ is two sets: the \emph{state variables} $x_i$ and the \emph{control variables} $C_i$:
\begin{empheq}[left={\empheqlbrace\,}]{align}\label{eq:control and states variables}
        X &= (x_1,x_2,x_3,...,x_s)\\
        C &= (C_1,C_2,C_3,...,C_c)
\end{empheq}

\noindent where $s$ is called the corank, and $c$ the codimension \cite{berry_iv_1980}. In practice, when dealing with physical systems, the state variables are the ones which cannot be controlled by an external stimuli, e.g. the pupil plane coordinates of a ray \cite{nye_orientations_1984} in optics. Conversely, control variables are the ones which can be adjusted by an external action, e.g. the choice of the image plane in the optical case \cite{nye_catastrophe_1986}. The definition of these two sets of variables associated to the correct generating function will be the first task to tackle in order to mathematically define the behaviour of the physical system, and thus properly apply catastrophe theory to the study of its discontinuities. \\
Then, the generating function is usually split in two parts \cite{barbero_catastrophe_2022} : 
\begin{equation}\label{eq:germ and unfolding term}
    \Phi (X,C) = g(X)+u(X,C)
\end{equation}

\noindent where $g$ is called the germ and $u$ is the unfolding term, where the control variables appear. The equilibrium surfaces can be determined by taking the derivatives of a generating function describing the dynamical behaviour of the system. Indeed, thanks to the variational principle, the equilibrium surface that results from the differential behaviour's characteristic function being minimized can always be used to describe the dynamical behaviour of any physical phenomenon. Hence, this equilibrium surface will simply correspond to the critical set of $\Phi$, which verifies : 

\begin{equation}\label{eq:critical set Phi}
    \frac{\partial\Phi}{\partial x_1} = \frac{\partial\Phi}{\partial x_2} = ... = \frac{\partial\Phi}{\partial x_s} = 0
\end{equation}

\noindent Equation \eqref{eq:critical set Phi} allows us to map the equilibrium surface inside a volume defined by the control and the states variables \cite{berry_waves_1976}. For instance if one has two state variables ($X \in \mathbf{R}^2$) and one control variable ($C\in \mathbf{R}$), the equilibrium surface will indeed be a bidimensional surface embedded in a volume $\mathbf{R}^2\times \mathbf{R}$ \cite{nye_relation_2005}.\\

In order to study the caustic as discontinuities in the physical behaviour of our system using the catastrophe theory, we have to compute a subset of the critical set of $\Phi$ called the degenerated critical set. To do so, let's consider the Hessian of $\Phi$: 
\begin{equation}\label{eq:Hessian}
    H[\Phi] \equiv \left(\frac{\partial^2\Phi}{\partial x_i\partial x_j}\right)_{i,j}
\end{equation}

\noindent Indeed, mathematically, the degenerated critical set of $\Phi$ corresponds to the positions which will satisfy \cite{appel_mathematics_2007}:
\begin{equation}\label{eq:det(H)=0}
    \text{det}(H[\Phi]) = 0
\end{equation}

\noindent In that context, caustics are singularities of gradient maps of the form \eqref{eq:critical set Phi} derived from generating functions $\Phi (X,C)$. Catastrophe theory - as developed around Thom's theorem - enables certain caustics, those which are \emph{structurally stable}, to be classified according to their topology \cite{thom_stabilite_1974,arnold_catastrophe_1992}. In other words, when the characteristic function of our system is perturbed, the local structure of the caustic remains unaltered because a smooth reversible transformation connects the perturbed and unperturbed caustics known as diffeomorphisms \cite{golubitsky_introduction_1978}. On the other hand, some non-generic scenario with unique symmetry can lead to a structurally unstable caustics, whose topological type is then altered by perturbations \cite{brocker_differentiable_1975}.\\

Thom's theorem states that for every value of the co-dimension $c$ of the control parameter space $C\in \mathbf{R}^c$, there are only a limited number of structurally stable caustics. In addition, for $c<5$, it provides seven explicit standard forms for the generating functions called the \emph{elementary catastrophes} \cite{thom_stabilite_1974,berry_iv_1980,deakin_elementary_1978}, classified in Table \ref{tab:elementary catastrophes}.

{\renewcommand{\arraystretch}{1.5}
\begin{table*}\
\centering
\begin{tabular}{|l|c|c|c|}
\hline
Name & c & s & $\Phi (X,C)$  \\\hline
Fold & 1 & 1 & $\frac{x^3}{3}+Cx$ \\\hline
Cusp & 2 & 1 & $\frac{x^4}{4}+C_2\frac{x^2}{2}+C_1x$ \\\hline
Swallowtail & 3 & 1 & $\frac{x^5}{5}+C_3\frac{x^3}{3}+C_2\frac{x^2}{2}+C_1x$\\\hline
Ellipic umbilic & 3 & 2 & $x_1^3-3x_1x_2^2+C_3(x_1^2+x_2^2)+C_2x_2+C_1x_1$\\\hline
Hyperbolique umbilic & 3 & 2 & $x_1^3+x_2^3+C_3x_1x_2+C_2x_2+C_1x_1$ \\\hline
Butterfly & 4 & 1 & $\frac{x^6}{6}+C_4\frac{x^4}{4}+C_3\frac{x^3}{3}+C_2\frac{s^2}{2}+C_1x_1$\\\hline
Parabolic umbilic & 4 & 2 & $x_1^4+x_1x_2^2+C_4x_2^2+C_3x_1^2+C_2x_2+C_1x_1$\\\hline
\end{tabular}
\caption{\label{tab:elementary catastrophes}The seven elementary catastrophes classification.}
\end{table*}}

Catastrophes characterized by $s = 1$ are called cupsoïd, while $s = 2$ leads to so-called umbilic catastrophes. Remarkably, adding more state variables $X$ has no effect on the structurally stable caustics for fixed $c$. It must be emphasized that - depending on the type of caustic - only one component $(x_1)$ or two $(x_1, x_2)$ of $X$ determine the caustic topology ; all the other components of $X$ can always be transformed to enter $\Phi (X,C)$ quadratically \cite{golubitsky_introduction_1978}, after which they cannot produce singularities when inserted inside equation \eqref{eq:critical set Phi}. \\

\subsection{First example: Euler's arch} \label{SEC:zeeman machine}

We would like to quickly illustrate this method with a situation familiar to any physicist. This mechanical analogy is inspired by Zeeman's groundbreaking work - which has contributed to the understanding of Thom's catastrophe theory - usually called the \emph{Zeeman's machine}, a simple mechanical scenario close to the optical situations studied in this work \cite{hilton_euler_1976,waddington_biological_2017,cazzolli_elastica_2020}.

\begin{figure}
\centering
\includegraphics[width=1.0\columnwidth]{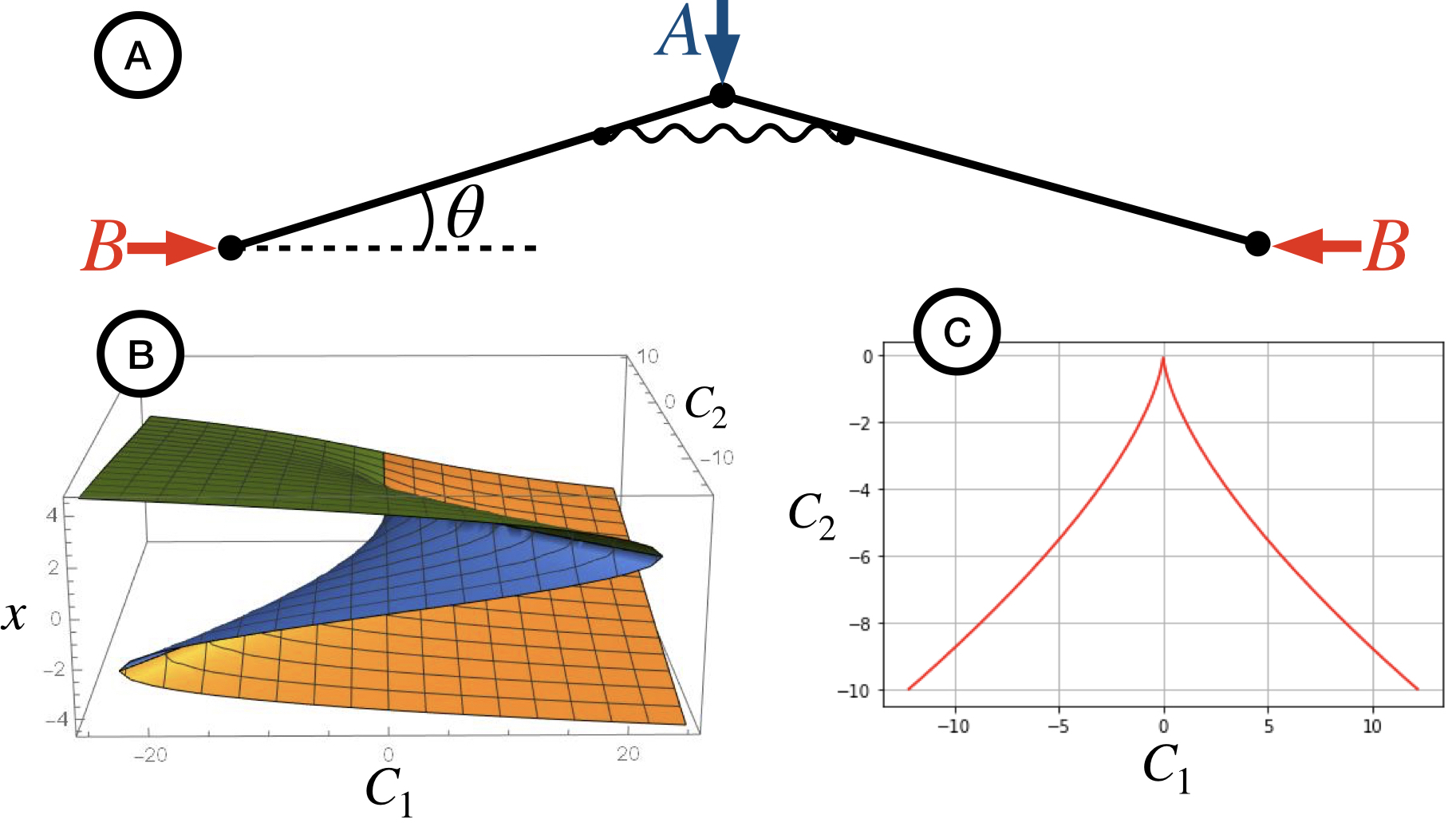}
\caption{A: An elastic hinge allowed the two rigid rods that made up the Euler arch to pivot. B force compresses the two ends. Consequently, the angle $\theta$ rises. The load called A will be also imposed. The arch gives way upwards (or downwards) at a critical number of these control parameters. B: Plot of the equilibrium positions of the arch using new control and state variables (see text). C: Projection of the equilibrium surface onto the control space depicting the fold catastrophe. }
\label{fig2}
\end{figure}

We begin by describing an Euler arch, as shown on Fig.~\ref{fig2}(a), which is made up of two rigid rods pivoting around a flexible hinge \cite{hilton_euler_1976,poston_catastrophe_1996}. Let's start by positionning the arch horizontally ($\theta=0$ on Fig.~\ref{fig2}(a)). We then apply a force $B$ at both ends of the arch. For an increasing value of B, the arch will bend upward (or downward). For the sake of the argument, let's assume that the arch buckles upwards, see Fig.~\ref{fig2}(a). We then apply a load $A$ on top of the arch. Slowly increasing $A$, one will eventually reach a critical value for which the arch abruptly snaps into the downward position. To investigate the reason for this behaviour, we need to consider the characteristic function of this problem - which corresponds to the elastic energy $V$ stored in the arch. The latter can be written using a fourth order Taylor expansion as \cite{poston_catastrophe_1996}: 
\begin{equation}\label{eulerarch}
 V=\frac{2\mu +\beta}{12}\theta^4-\frac{a}{6}\theta^3-B\theta^2+A\theta
\end{equation}

\noindent where $\beta=B-2\mu$ and $\mu$ is the elastic constant of the spring. Using the catastrophe theory framework we can define the state variable $x=\theta -\frac{a}{2(2\mu+B)}$ and the two new constants $C_1$ and $C_2$ - composed using the physical parameters of our system $A,B,\beta,\mu$ - that will be considered as the two control variables of our problem. With those definitions, it is now possible to write the characteristic function according to the cusp generating function - corresponding to one of the seven elementary catastrophe (see Table (\ref{tab:elementary catastrophes})) :
\begin{equation}\label{eq:function_cusp}
    \Phi (x,C_2,C_1) = \frac{x^4}{4}+C_2\frac{x^2}{2}+C_1x
\end{equation}

\noindent Then, the arch's equilibrium positions are given by \eqref{eq:critical set Phi}: 
\begin{equation}\label{eq:critical set cusp}
    x^3+C_2x+C_1 = 0
\end{equation}

\noindent This cubic equation for $x$ has one or three roots, depending on $C_1$ and $C_2$. This equilibrium condition can be mapped into the space $(x,C_1,C_2)$ and corresponds to a folded surface (Fig.~\ref{fig2}(b)), usually called the \emph{catastrophe manifold} \cite{zeeman_catastrophe_1976}. Eventually, the degenerated critical set of $\Phi$ can be found by applying equation \eqref{eq:det(H)=0}: 
\begin{eqnarray}
    3x^2+C_2 &=& 0 \label{eq:degenerated set cusp1}\\
    \implies x &=& \pm \sqrt{\frac{-C_2}{3}}\label{eq:degenerated set cusp2}
\end{eqnarray}

\noindent To investigate abrupt discontinuities, we can locate the gradient map singularities by using expressions (\ref{eq:degenerated set cusp2}). Indeed, they will map in the control parameter space their values causing sudden changes in the system's state parameters. The projection of these positions located on the equilibrium surface onto the control space are called \emph{bifurcation set} \cite{zeeman_catastrophe_1976}. Thom's theorem states that if the generating function is expressed as \eqref{eq:function_cusp}, then the bifurcation set will always be a cusp, as shown in Fig.~\ref{fig2}(c), no matter the exact experimental conditions encountered in the laboratory.
Hence, by combining equations (\ref{eq:critical set cusp}) and (\ref{eq:degenerated set cusp2}) we find the parametrization for the cusp :
\begin{equation}\label{eq:parametrization cusp}
    (x,C_2,C_1) = (x,-3x^2,2x^3)
\end{equation}

\noindent By eliminating the state variable $x$ in \eqref{eq:parametrization cusp}, one can also compute the equation of the bifurcation set directly in the control space, expressed as a semicubical parabola with a cusp at the origin :
\begin{equation}\label{cuspincontrolspace}
4C_2^3+27C_1^2=0
\end{equation}

\noindent Fig.~\ref{fig2}(c) shows the cusp in control space, for $C_1\in[ -10,10]$ and $C_2\in[ -10,0]$.\\

The previous example aimed at illustrating the overall procedure in the scenario of the most well-known catastrophe defined by a codimension 1 and a corank 2, known as the cusp. Higher codimensions catastrophes will be the focus of our further investigation in the context of charged particle optics. However, their structurally stable caustics will be interpreted using the same mathematical approach.

\subsection{Second example: Catastrophe optics}

The specific area of optics where Thom's theory is applied is known as \emph{catastrophe optics}. It was first extensively studied by M. Berry and J. Nye in the context of conventional optics \cite{berry_iv_1980,nye_natural_1999}. The purpose of our work is to extend these works to the field of charged particles optics, for applications in electron and ion microscopy. As a preamble, let us briefly review Berry and Nye's equivalent methodology within the framework of standard optics, as given in their corresponding reference works \cite{berry_iv_1980,nye_natural_1999,berry_waves_1976,nye_relation_2005}.\\ 

\begin{figure}
\centering
\includegraphics[width=\columnwidth]{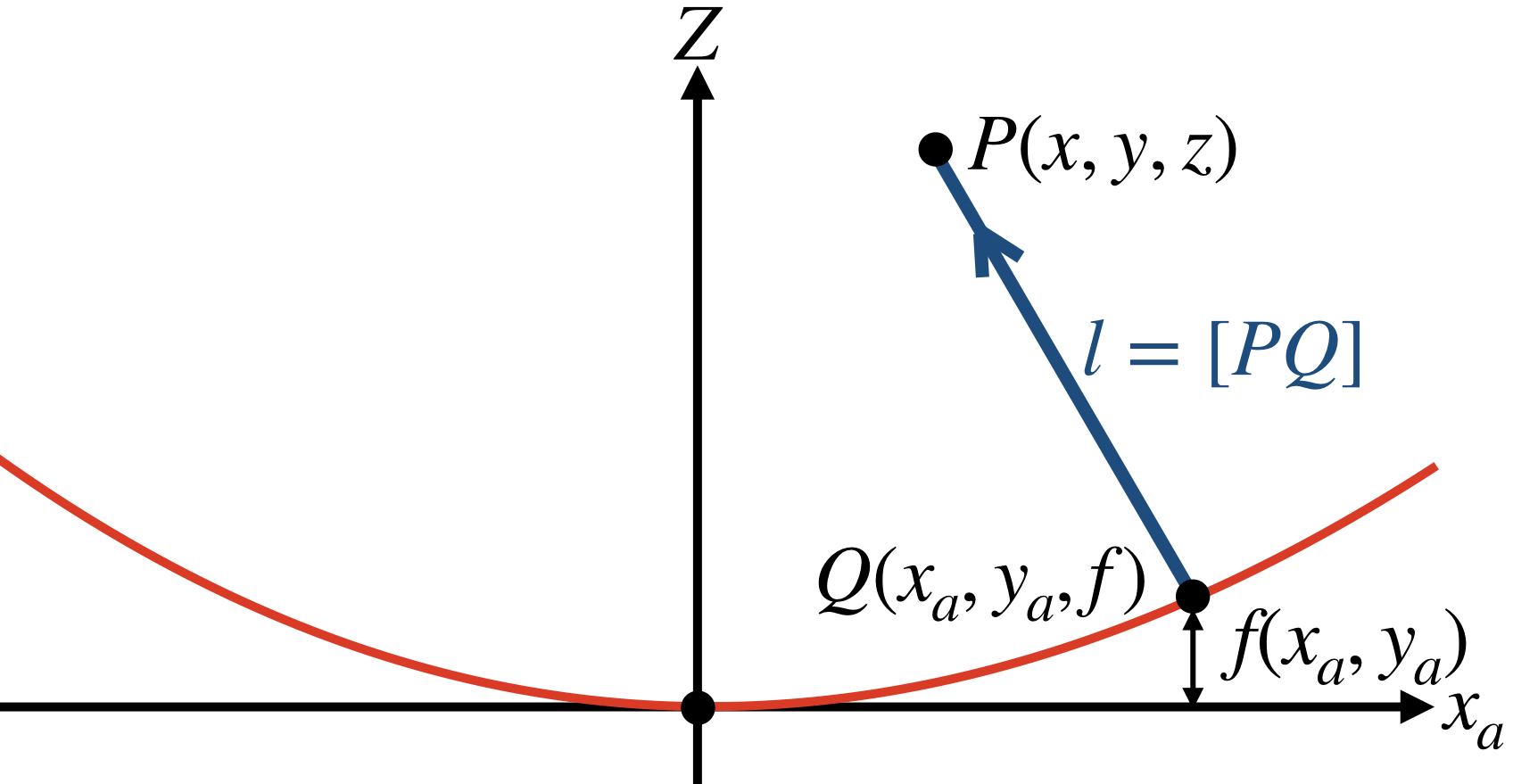}
\caption{Optical path $l$ between the exit pupil plane of the system, located in $z=0$, and the observation plane located at the coordinate $Z$.}
\label{fig3}
\end{figure}

First of all the pupil plane will be chosen as the state space for our generating function, while the control space will select the position of the observation plane. Following the Fermat's principle and the work of Hamilton, the point characteristic function will be the optical path. It is selected to be the generating function $\Phi$ on which the catastrophe theory will be applied. First, this generating function must be expressed in the same structure as one of the seven elementary catastrophes listed in table (\ref{tab:elementary catastrophes}). To do so, we will use the same formalism as the one described by J.F. Nye in \cite{nye_symmetrical_2018,nye_natural_1999}. A simple representation of the image side of an optical system is shown on figure \ref{fig3}, where the optical axis is represented by the $Z$-axis, and its origin is taken at the exit pupil plane.

We are looking for the optical path $l$ between $P$ and $Q$. $Q$ is a point taken in the wavefront, with coordinate $(x_a,y_a,f)$, where $f(x_a,y_a)$ is the distance between the pupil plane and the wavefront at the coordinates $(x_a,y_a)$. In the following, we are going to explore the observation as a function of $Z$. Points $P(x,y,z)$ which are part of the degenerated critical set of $\Phi$ will be called the caustic points. Since we assume that there is no lens between $P$ and $Q$, the optical path between these two locations corresponds to a straight line expressed as follow : 
\begin{equation}\label{eq:straght line CCO}
    l = \sqrt{(x-x_a)^2+(y-y_a)^2+(z-f)^2}
\end{equation}

\noindent We now consider $Z\gg f$ and expand the square root using the the far-field approximation, which implies that $Z \gg X-x,Y-y$. Thus, equation \eqref{eq:straght line CCO} can be written at first order as:
\begin{equation}\label{eq:far-field CCO}
    l \approx z-f+\frac{(x-x_a)^2}{2z}+\frac{(y-y_a)^2}{2z}
\end{equation}

\noindent In this case, the equilibrium conditions \eqref{eq:critical set Phi} can be found using the Fermat's principle \eqref{fermatderivationcaustic}. After using this principle, the first term in equation \eqref{eq:far-field CCO} will vanish and not contribute to the equilibrium surface. We can then define the generating function $\Phi$ - sometimes called Fermat's potential in the literature - as follow \cite{nye_symmetrical_2018,barbero_catastrophe_2022}:

\begin{eqnarray}
        \Phi &=& z+\frac{x^2+y^2}{2z}-l \label{eq:fermat potential CCO 1}\\
         &=& f - \frac{x_a^2+y_a^2}{2z}+\frac{x_ax+y_ay}{z} \label{eq:fermat potential CCO 2}    
\end{eqnarray}

\noindent Then, we admit the following relation that comes from \cite{nye_symmetrical_2018}: 

\begin{equation}\label{eq:wavefront aberrations in f}
    f(x_a,y_a) = W(x_a,y_a)+\frac{x_a^2+y_a^2}{2Z_0}
\end{equation}

\noindent where $Z_0$ is the distance between the paraxial image plane and the pupil plane, and $W$ represents the wavefront aberrations. Inserting \eqref{eq:wavefront aberrations in f} in \eqref{eq:fermat potential CCO 2}, we obtain the final expression of the generating function :
\begin{equation}
    \Phi = W + \left(\frac{1}{2Z_0}-\frac{1}{2z}\right)\left(x_a^2+y_a^2\right)+\frac{x_ax}{z}+\frac{y_ay}{z}
\end{equation}

\noindent Thom's theorem then requires that we express $\Phi$ in a canonical form. To do so, one can then define the following three control variables: 
\begin{empheq}[left={\empheqlbrace\,}]{align}\label{eq:control variables OCC}
        C_1 &= \frac{x}{z}\\
        C_2 &= \frac{y}{z}\\
        C_3 &= \frac{1}{2}\left(\frac{1}{Z_0}-\frac{1}{z}\right)
\end{empheq}

\noindent Besides, as was previously mentioned, the pupil coordinates $x_a$ and $y_a$ correspond to the state variables. The canonical generating function therefore corresponds to: 
\begin{equation}\label{eq:PHI OCC}
    \begin{split}
    \Phi(x_a,y_a,C_3,C_2,C_1) = W(x_a,y_a) &+C_3(x_a^2+y_a^2)\\
    &+C_2y_a+C_1x_a
    \end{split}
\end{equation}

\noindent The critical set of $\Phi$ is found by applying \eqref{eq:critical set Phi}:
\begin{equation}\label{eq:critical set of Phi CCO}
    \frac{\partial\Phi}{\partial x_a} = \frac{\partial\Phi}{\partial y_a} = 0
\end{equation}

\noindent The caustics observed in the plane at position $z$ correspond to the regions $(x,y)$ in which a high intensity is created by many rays going through the same point, but coming from different initial pupil positions. According to the catastrophic optics vocabulary, the caustics correspond to the points map in the control space that coincide with the degenerated critical set of the state space \cite{nye_symmetrical_2018,barbero_catastrophe_2022}. This set verifies equation \eqref{eq:det(H)=0}, and therefore the following equation : 
\begin{equation}\label{eq:degenerated set CCO}
    \frac{\partial^2\Phi}{\partial x_a^2}\frac{\partial^2\Phi}{\partial y_a^2}-\left(\frac{\partial^2\Phi}{\partial x_a\partial y_a} \right)^2= 0
\end{equation}

\noindent Using this method, caustic surfaces are be obtain in one go, as sets of points that follow (\ref{eq:degenerated set CCO}) and (\ref{eq:critical set of Phi CCO}), rather than through the computation of each rays' trajectory. This strategy is strictly equivalent to the one described in \ref{SEC:zeeman machine}, when replacing the elastic energy by the optical path. In addition, depending on the expression of the wavefront $W$ in \eqref{eq:PHI OCC}, the caustic will follow the rules of one of the seven Thom's elementary catastrophes. If a complex wavefront yields a generating function with codimensions and corank outside the validity range of Thom's framework, one needs to follow Arnold's approach, which extends Thom's results to more complex topological generating functions \cite{arnold_theory_1990}. \\

The primary objective of this introduction was to illustrate the efficiency and generality of the catastrophe theory. Moreover, we have shown why the application of catastrophe theory to conventional optics should be grounded in Hamilton's theory of characteristic function and Fermat's principle. For charged particle optical systems, the equivalent will be obtained thanks to a modified Lagrangian function and the least action principle. 

\section{Charged particles optics}\label{sec:OPC}

\subsection{General approach}

As previously introduced, Hamilton demonstrated how the laws of mechanics and optics can be derived by using the mathematical properties of a single characteristic function \cite{hamilton_general_1833}. In that context, charged-particle optics (CPO) could also be treated using this powerful method of point, mixed and angular characteristic functions \cite{rose_geometrical_2009}, which generally depend on six parameters - e.g. the position coordinates of a ray at the object and image areas for the point characteristic functions, or the angular coordinates of a ray at the object and image spaces for the angular characteristic functions. 
Nevertheless, since an arbitrary trajectory in space is fully characterized by four parameters, Heinrich Bruns demonstrated in 1895 that it is actually preferable to give the characteristic of an optical system as a function of four variables \cite{bruns_eikonal_1895}. These modified characteristic functions are generally called \emph{eikonals}. 
Similar to conventional optics, perturbation methods are usually used to determine the eikonals \cite{hawkes_principles_2018,sturrock_perturbation_1951}. These methods, which are based on the perturbation characteristic function approach, provide a useful substitute for the widely used trajectory method, which determine the Lorentz force action on a charged particle by solving the Newton's equation of motion \cite{glaser_zur_1935}. \\

In the following, we will always assume that the particles travel only in regions which are free of space currents and charges. They will interact with static fields (either electrostatic or magnetostatic) which can be derived from potentials obeying the Poisson equation. Using analytical mechanics we will derive an equivalent of the optical index encountered in conventional optics. Then Hamilton's description of the least action principle will provide us with an equivalent of Fermat's principle, which states that the system minimizes the "optical path" of the charged particles. Later, this definition of the optical path will be used as the generating function to apply Thom's catastrophe theory and Fermat's principle will give us an equilibrium surface \cite{thom_stabilite_1974,buchdahl_introduction_1970}. The singularities of this equilibrium surface will then be studied to understand the geometry of the caustics.\\

\subsection{Hamiltonian picture of charged particles optics}

In this section, we will define our generating function by going over the charged particle optics formalism. A dense literature have gone into great detail about this formalism \cite{hawkes_principles_2018,rose_geometrical_2009,szilagyi_electron_1988} which will be briefly reviewed in the following. In particular, we have chosen to use H. Rose's formalism as our foundation for the next parts, as charged particle optics lacks a standard nomenclature. In this section, a condensed outline of the essential informations needed to fully grasp our application will be presented, while a detailed exposition can be found in \cite{rose_geometrical_2009}.\\

The eikonal approach is based on the analytical mechanics formalism. It is derived using the Hamiltonian $H$ and the Lagrangian $\mathcal{L}$ functions defined using particle kinetic $K$ and potential energies $V$ (of a charged particle inside static fields in our case). In that case the Lagrangian is given simply by the difference between these two energies $  \mathcal{L} = K-V$, while the Hamiltonian is the sum of these two energies $H = K+V$ \cite{lanczos_variational_1986}.
The action - which is then defined as the integral of the Lagrangian over time - is used to apply the principle of least action \cite{cline_variational_2019}. \\
The Lagrangian of a charged particle (an electron in that case) inside an electromagnetic field is explicitly  given  by \cite{rose_geometrical_2009} :

\begin{equation}\label{eq:lagrangian}
    \mathcal{L}=  m(v^2-c^2)-e\Vec{v}.\Vec{A}+e\phi
\end{equation}

\noindent where $e$, $m$ and $\vec{v}$ are respectively the charge, the mass \footnote{$m$ is the relativistic mass $m = \gamma m_e$, $\gamma$ being the Lorentz factor, $m_e$ the mass of the electron at rest.} and the speed of the electron. $\phi$ denotes the electrostatic potential and $\vec{A}$ is the magnetostatic vector potential, both generated by electromagnetic lenses. Using the Lagrangian, we can define the action $W$ \cite{cline_variational_2019} as: 
\begin{equation}\label{eq:action}
    W = \int_{t_0}^t  \mathcal{L}(\vec{r},\dot{\vec{r}},t)dt
\end{equation}

\noindent where we assumed that the particle goes from $\vec{r}_0$ at $t_0$ to the point $\vec{r}$ at $t$. Using Legendre transformation, one can write: $L = \vec{p}.\Dot{\vec{r}}-H$, with $\dot{\vec{r}} = d\vec{r}/dt$. Since, we will discuss only stationary optical system, we obtain:
\begin{eqnarray}
        S &=& W+E(t-t_0)\\ 
        &=& \int_{t_0}^t \vec{p}.\dot{\vec{r}}dt-\int_{t_0}^tHdt+E(t-t_0)\\
        &=& \int_{\vec{r}_0}^{\vec{r}}\vec{p}\cdot d\vec{r}
\end{eqnarray}

\noindent with $E$ corresponding to the energy of the system and $S$ the reduced action - or eikonal \cite{rose_geometrical_2009}. Because the optical axis corresponds to the $z$ axis one can define :
\begin{eqnarray}\label{eq:def reduced action}
        \tilde{  \mathcal{L}} &=& \vec{p}\frac{d\vec{r}}{dz}\\
        \implies S &=& \int_{z_0}^z \tilde{\mathcal{L}} \; dz
\end{eqnarray}

\noindent Such as Fermat's principle in conventional optics, the Maupertuis principle in variational mechanics states that \cite{cline_variational_2019}: 
\begin{equation}\label{eq:maupertuis principle}
    \delta S = 0
\end{equation}

\noindent Because we are dealing with electromagnetic systems, it will be more convenient to work using complex notations by defining $w = x+iy$, $ \overline{w} = x-iy$, $w'= \frac{dw}{dz}$. We moreover set $w_0 = w(z_0)$ since $w$ is a function of $z$ i.e. $w = w(z)$. The equations of motion can be expressed using Euler-Lagrange (EL) formulation of \eqref{eq:maupertuis principle} which gives \cite{cline_variational_2019} : 

\begin{empheq}[left={\empheqlbrace\,}]{align}
         \frac{d}{dz}\left(\frac{\partial\Tilde{  \mathcal{L}}}{\partial\overline{w}'} \right) &= \frac{\partial\Tilde{  \mathcal{L}}}{\partial\overline{w}} \label{eq:motion L tilde 1}\\
         \frac{d}{dz}\left(\frac{\partial\Tilde{  \mathcal{L}}}{\partial w'} \right) &= \frac{\partial\Tilde{  \mathcal{L}}}{\partial w} \label{eq:motion L tilde 2}
\end{empheq}

\noindent Expressing \eqref{eq:def reduced action} using the complex variables one can show:
\begin{eqnarray}
        \tilde{  \mathcal{L}} &=& \vec{p}\frac{d\vec{r}}{dz}\\
        &=& p_xx'+p_yy'+p_z\\
        &=& \frac{1}{2}(p\overline{w}'+\overline{p}w')+p_z
\end{eqnarray}

\noindent which leads to:
\begin{equation}\label{eq: p}
     \frac{\partial\tilde{  \mathcal{L}}}{\partial\overline{w}'} = \frac{1}{2}p 
\end{equation}

\noindent where $p = p_x+ip_y$ corresponds to the complex lateral component of the canonical momentum. The variational function $\delta S$ can be expressed through an integration by part of \eqref{eq:def reduced action}. Then, using EL equations \eqref{eq:motion L tilde 1} and \eqref{eq:motion L tilde 2}, together with \eqref{eq: p}, we eventually obtain the following expression \cite{rose_hamiltonian_1987} :
\begin{equation}
\label{eikonal1}
    \delta S = \Re(p\delta\overline{w}-p_0\delta\overline{w}_0)
\end{equation}

\noindent leading to the well-known conventional optics equations enabling to express the values of one set of variables (momentum or space variables) with the derivate of the eikonal function relatively to the other set (in object or image space) \cite{synge_geometrical_1937,buchdahl_introduction_1993,luneburg_mathematical_2021}. $S$ is a function of space coordinates $w_0$, $\overline{w}_0$, $w$ and $\overline{w}$ in object and image spaces. It then corresponds to the point eikonal which is usually called Hamilton's point characteristic function in conventional optics. Hence, thanks to \eqref{eikonal1}, we can express the momentum coordinates through a simple differentiation of the point eikonal function: 
\begin{empheq}[left={\empheqlbrace\,}]{align}
        p_0 &= -2\frac{\partial S}{\partial \overline{w}_0}\label{canonicalpointeikonal 1}\\
        p &= 2\frac{\partial S}{\partial\overline{w}}\label{canonicalpointeikonal 2}
\end{empheq}

\noindent Other eikonal functions can be expressed from the Legendre transformation of $S$ using (\ref{canonicalpointeikonal 1}). For instance the angle-point eikonal function - which depends on the momentum $p_0$ and $\overline{p}_0$ in the object space, and the space coordinates $w$ and $\overline{w}$ in the image space - can be expressed as: \cite{luneburg_mathematical_2021,rose_geometrical_2009}: 

\begin{empheq}[left={\empheqlbrace\,}]{align}
        V(p_0,\overline{p}_0,w,\overline{w}) &= S+\Re(p_0w_0)\\
        \delta V &= \Re(p\delta\overline{w}+\overline{w}_0\delta p_0)\label{eq:legendre angle point eikonal}
\end{empheq}

In order to express these eikonal functions as a function of electric and magnetic potentials along the charged particles trajectories, we will first define the energy normalized eikonal function $L = S/q_0$, with $q_0 = \sqrt{2em_e\Phi_0^*}$ the normalized momentum linked to the acceleration voltage of the charged particles. $\Phi_0$ is then the potential on the optical axis at $z_0$, and $*$ is used to specify when the relativistic correction has to be taken into account. Like for the Lagrangian in \eqref{eq:def reduced action}, we can also define a new variational function $\mu$ using the normalized eikonal function as:  
\begin{eqnarray}
    \label{eq:variationnal function mu}
    L &=& \frac{S}{q_0} = \int_{z_0}^z\mu dz\\
    \implies \mu &=&\frac{1}{q_0}\vec{p}\frac{d\vec{r}}{dz}
\end{eqnarray}

\noindent By comparison with Fermat's principle, this variational function $\mu$ then corresponds to the classical definition of the optical index in conventional optic - but applied to CPO systems \cite{buchdahl_introduction_1970,torre_linear_2005}. Here, the optical index in CPO doesn't change abruptly at an interface, but can be regarded as the equivalent of a gradient of optical index in conventional optics \cite{rose_geometrical_2009}. Using the expression of the canonical momentum of a charged particle inside an electromagnetic field $\vec{p} = m\vec{v}-e\vec{A}$, one can split the variational function in two optical indices: the electrostatic $\mu_e$ and magnetostatic $\mu_m $ contributions to the general variational function $\mu$. These two functions can be expressed as follow:
\begin{eqnarray}
        \mu_e &=& \sqrt{\frac{\phi^*}{\Phi_0^*}(w'\overline{w}'+1)}\label{eq:mu_m and mu_e result 1} \\
        \mu_m &=& -\frac{e}{q_0}(A_z+\Re(\overline{A}w')) \label{eq:mu_m and mu_e result 2}
 \end{eqnarray}
 
\noindent where $A = A_x+iA_y$ corresponds to the complex radial component of $\vec{A}$. We have define the relativistically-corrected potentials following H. Rose's notations \cite{rose_geometrical_2009}: $m^2v^2 = 2em\phi^*$, with $v = |\vec{v}|$ and $\phi^* = \phi(1+e\phi/2m_ec^2)$.\\

\subsection{Polynomial expansion}

In order to extract the paraxial and aberrations' contributions to the charged particles trajectories, we have to perform a polynomial expansion of the two variational functions \eqref{eq:mu_m and mu_e result 1} and \eqref{eq:mu_m and mu_e result 2} with respect to space and momentum variables.

Let's first consider the electrostatic scalar potential $\phi$. This potential can be developed using an azimuthal Fourier series in any plane perpendicular to the optical axis (i.e. the $z$ axis) as: 
\begin{equation}\label{eq:TF of phi}
    \phi = \sum_{\nu = 0}^\infty\phi_\nu
\end{equation}

\noindent where $\phi_\nu$ are the harmonic potentials given by the general expression :
\begin{equation}\label{eq:phi_nu}
    \phi_\nu = \Re\sum_{\lambda = 0}^\infty a_{\nu\lambda}(z)(w\overline{w})^\lambda\overline{w}^\nu
\end{equation}

\noindent Since each harmonic potential \eqref{eq:phi_nu} is a solution of the Laplace equation, one can show that the components $a_{\nu\lambda}(z)$ can be expressed as a function of the potential extracted from the optical axis $\Phi(z) = \phi(x=y=0,z)$ and its derivatives \cite{rose_geometrical_2009}. One can define the harmonic components $a_{\nu 0}(z) = \Phi_\nu(z) $, and insert \eqref{eq:phi_nu} inside the Laplace equation. After few mathematical manipulations, one can show that any component $ a_{\nu\lambda}$ can be expressed as \cite{rose_geometrical_2009}:
\begin{equation}\label{eq:value of a_nulambda}
    a_{\nu\lambda} = (-)^\lambda\frac{1}{4^\lambda\lambda!}\frac{\nu!}{(\lambda+\nu)!}\Phi_\nu^ {[2\lambda]}
\end{equation}

\noindent where $\Phi_\nu^{[i]}$ corresponds to the $i^{th}$ derivative of $\Phi_\nu$. Then, injecting these expressions into equation \eqref{eq:phi_nu}, we obtain the general expression of the potential expansion: 
\begin{equation} \label{eq:phi expension 1}
        \phi_\nu = \sum_{\lambda = 0}^\infty(-)^\lambda\frac{\nu!}{\lambda!(\lambda+\nu)!}\left(\frac{w\overline{w}}{4}\right)^\lambda \Re(\Phi_\nu^ {[2\lambda]}(z)\overline{w}^\nu)
\end{equation}

\noindent and:
\begin{equation} \label{eq:phi expension 2}
    \phi = \sum_{\nu = 0}^\infty\sum_{\lambda = 0}^\infty(-)^\lambda\frac{\nu!}{\lambda!(\lambda+\nu)!}\left(\frac{w\overline{w}}{4}\right)^\lambda \Re(\Phi_\nu^ {[2\lambda]}(z)\overline{w}^\nu)
\end{equation}

The same operation can be implemented for magnetostatic systems. The static magnetic flux density vector inside a magnetic system is given by $\Vec{B} = -\vec{\nabla}\psi = \vec{\nabla}\wedge\Vec{A} $. Indeed, the scalar magnetic potential $\psi$ can be used instead of $\Vec{A}$ for non-saturated magnetic component \cite{yavor_optics_2009}. Following H. Rose, we define the conjugate part of $A$ noted $ \overline{A}  = A_x-iA_y$ and the $z$ component of the magnetic vector potential $A_z$. We express them also using a polynomial expansion of each harmonic components given by the same azimuthal Fourier decomposition procedure as before (see \cite{rose_geometrical_2009} for details). We find:
\begin{equation}\label{eq:A expension 1}
    A_z = \sum_{\nu = 0}^\infty\sum_{\lambda=0}^\infty(-)^\lambda\frac{\nu!}{\lambda!(\lambda+\nu)!}\left(\frac{w\overline{w}}{4}\right)^\lambda \Im(\Psi_\nu^{[2\lambda]}(z)\overline{w}^\nu)
\end{equation}

\noindent and:
\begin{equation}\label{eq:A expension 2}
    \overline{A} = \frac{i}{2}\overline{w}\sum_{\nu = 0}^\infty\sum_{\lambda=0}^\infty(-)^\lambda\frac{\nu!}{\lambda!(\lambda+\nu+1)!}\left(\frac{w\overline{w}}{4}\right)^\lambda\Psi_\nu^{[2\lambda+1]}(z)\overline{w}^\nu
\end{equation}

\noindent here $\Psi_\nu$ is the analogue of $\Phi_\nu$ - as it corresponds to the value of the scalar magnetic potential along the optic axis $\psi_\nu(x=0,y=0,z)=\Psi(z)$, and $\psi = \sum_{\nu = 0}^\infty\psi_\nu$.\\

In summary, \eqref{eq:phi expension 1}, \eqref{eq:phi expension 2}, \eqref{eq:A expension 1} and \eqref{eq:A expension 2} are simply polynomial expansions where the variables of each polynomial are $w$, $\overline{w}$, $w'$ and $\overline{w}'$. Electron microscopes' optical elements can be regarded as one (or a superposition of different) electrostatic or magnetostatic harmonic components (which could be either complex or real functions) among the five main symmetries given in the following list :
\begin{itemize}
    \item $\nu = 0$ : Round lens (and acceleration voltage)
    \item $\nu = 1$ : Dipolar lens
    \item $\nu = 2$ : Quadrupolar lens (used stigmator)
    \item $\nu = 3$ : Hexapolar lens
    \item $\nu = 4$ : Octopolar lens
\end{itemize}

\noindent When the polynomial expansions \eqref{eq:phi expension 1}, \eqref{eq:phi expension 2}, \eqref{eq:A expension 1} and \eqref{eq:A expension 2} are inserted in \eqref{eq:mu_m and mu_e result 1} and \eqref{eq:mu_m and mu_e result 2}, one can write the variational function $\mu$ as a polynomial expansion: 
\begin{equation}
    \mu = \sum_{n=0}^\infty\mu^{(n)}
\end{equation}

\noindent where $\mu^{(n)}$ correspond to a polynomial function of degree $n$ - related to the polynomial expansions of the variables $w$, $\overline{w}$, $w'$ and $\overline{w}'$. Within this notation system, the paraxial solutions of a CPO system can be extracted from this expansion using only the components $n\leq2$. 

\subsection {Paraxial solutions of general optical system}

We start by considering the three first terms of the polynomial expansion $\mu=\mu^{(0)}+\mu^{(1)}+\mu^{(2)}$. Interested readers may find in the literature an extensive amount of mathematical details related to these developments \cite{rose_geometrical_2009,hawkes_principles_2018,grivet_electron_1972}. Let's start with the expansion of  the magnetic component  $\mu_m$. We first have to extract from $A_z$ and $\overline{A}$ all the series which will appear in the polynomial expansion of order 2. According to \eqref{eq:A expension 1} and \eqref{eq:A expension 2}, each polynomial will be defined by terms of power $2\lambda+\nu$ - noted $A_z^{\nu\lambda}$. Hence, within the paraxial solutions, only the terms $A_z^{00}$,$A_z^{10}$,$A_z^{20}$ and $A_z^{01}$ have to be considered :
\begin{align}
        A_z^{00} &= \Im(\Psi_0(z)) = 0 \notag \\
        A_z^{01} &= -\frac{w\overline{w}}{4}\Im(\Psi_0''(z))= 0 \label{eq:Az expension for nleq2} \\
        A_z^{10} &= \Im(\Psi_1(z)\overline{w}) \notag \\
        A_z^{20} &= \Im(\Psi_2(z)\overline{w}^2) \notag 
\end{align}

\noindent The same procedure has to be implemented for $\overline{A}$. Only the term $\overline{A}^{00}$ among all $\overline{A}^{\nu\lambda}$ terms of the polynomial expansion of $\overline{A}$ appear in the paraxial expansion of $\mu$ :
\begin{align}
        \overline{A}^{00} &= \frac{i}{2}\overline{w}\Psi_0'(z)\notag\\
        \implies Re(\overline{A}^{00}w') &= \Re\left(\frac{i}{2}\overline{w}\Psi_0'(z)w'\right)\label{eq:Abarre expension for nleq1}\\
        \implies Re(\overline{A}^{00}w') &= -\Im\left(\frac{1}{2}\Psi_0'(z)\overline{w}w'\right)\notag
\end{align}

\noindent One can then insert equation (\ref{eq:Az expension for nleq2}) and (\ref{eq:Abarre expension for nleq1}) into relation (\ref{eq:mu_m and mu_e result 2}) to find: 
\begin{eqnarray}
        \mu_m^{(0)} &=& 0 \label{eq:mu_m^n parax 0} \\
        \mu_m^{(1)} &=& -\frac{e}{q_0}\Im(\Psi_1\overline{w})\label{eq:mu_m^n parax 1}\\
        \mu_m^{(2)} &=& \frac{e}{q_0}\Im\left(\frac{1}{2}\Psi_0'(z)\overline{w}w'-\Psi_2(z)\overline{w}^2 \right) \label{eq:mu_m^n parax 2}
\end{eqnarray}

\noindent The same operation can be performed for the electrostatic part $\mu_e$ and leads to \cite {rose_geometrical_2009}: 
\begin{align}
        \mu_e^{(0)} &= \sqrt{\frac{\Phi_0^*(z)}{\Phi_0^*(z_0)}} \label{eq:mu_e parax 0}\\
        \mu_e^{(1)} &= \frac{\gamma_0}{2\sqrt{\Phi_0^*(z_0)\Phi^*_0(z)}}\Re\left(\Phi_1(z)\overline{w}\right) \label{eq:mu_e parax 1}\\
        \begin{split}
        \mu_e^{(2)} &= \frac{1}{2}\sqrt{\frac{\Phi_0^*(z)}{\Phi_0^*(z_0)}}\Re\Bigg[ w'\overline{w'} -\frac{w\overline{w}}{\Phi_0^*(z)}\Bigg(\frac{\gamma_0\Phi_0''(z)}{4}\\
        &+\frac{\Phi_1(z)\overline{\Phi}_1(z)}{8\Phi_0^*(z)}\Bigg)
        +\frac{\overline{w}^2}{\Phi_0^*(z)}\Bigg(\gamma_0\Phi_2(z)-\frac{\Phi_1^2(z)}{8\Phi_0^*(z)}\Bigg) \Bigg]
        \end{split} \label{eq:mu_e parax 2}
\end{align}

\noindent with:
\begin{eqnarray}
        \epsilon &=& \frac{e}{2m_ec^2} \label{eq:epsilon and gamma_0 1}\\
        \gamma_0 &=& 1+2\epsilon\Phi_0(z) \label{eq:epsilon and gamma_0 2}\\
        v_0 &=& \frac{1}{\gamma_0}\sqrt{\frac{2e\Phi_0^*(z)}{m_e}} \label{eq:epsilon and gamma_0 2}\\
        \implies\phi^* &=& \phi+\epsilon\phi^2 \label{eq:epsilon and gamma_0 4}
\end{eqnarray}
\noindent where $\gamma_0$ is the Lorentz factor and $v_0$ is the velocity associated to the accelerated potential $\Phi_0$. Paraxial solutions can be found by inserting the $\mu$ polynomial expansion inside the EL equations \cite{cline_variational_2019} which leads to:
\begin{equation}\label{eq:mu in EL equation}
    \frac{d}{dz}\frac{\partial\mu}{\partial\overline{w}'}-\frac{\partial\mu}{\partial\overline{w}}=0
\end{equation}

\noindent From this operation, we can extract the paraxial differential equation of a general charged particles systems :
\begin{equation}\label{eq:equation paraxial}
\begin{split}
        &w''+\frac{\gamma_0}{2\Phi_0^*}(\Phi_0'+iv_0\Psi_0')w'\\
        &+\frac{\gamma_0}{4\Phi_0^*}\left(\Phi_0''+iv_0\Psi''_0+\frac{\Phi_1\overline{\Phi}_1}{2\gamma_0\Phi_0^*} \right)w\\
        &-\frac{\gamma_0}{\Phi_0^*}\left( \Phi_2+iv_0\Psi_2-\frac{\Phi^2_1}{8\gamma_0\Phi_0^*}\right)\overline{w}\\
        &= \frac{\gamma_0}{2\phi_0^*}(\Phi_1+iv_0\Psi_1)
\end{split}
\end{equation}

\noindent This equation is a second order partial differential equation with $z$-dependent coefficients. There is therefore no general analytical solution, and one has to solve it using adapted numerical integrations. We can nevertheless define four independent solutions $w_\mu, \mu = 1,2,3,4$  from the following boundary conditions \cite{rose_geometrical_2009} :

\begin{equation}\label{eq:general paraxial fundamental rays}
    \begin{gathered}
        w_1(z_0) = 1,w'_1(z_0) = 0\\
        w_2(z_0) = i,w'_2(z_0) = 0\\
        w_3(z_0) = 0,w'_3(z_0) = 1\\
        w_4(z_0) = 0,w'_4(z_0) = i
    \end{gathered}
\end{equation}

\noindent It is then always possible to express any skew rays as a linear combination of these four solutions:
\begin{equation}
    w(z) = x_0w_1(z)+y_0w_2(z)+x_0'w_3(z)+y_0'w_4(z)
\end{equation}

\noindent In that context, the principal ray corresponds to $w(z_0) = 1$ or $i$, $w'(z_0) = 0$ boundary conditions while the marginal ray is defined by the second set of boundary conditions $w'(z_0) = 1$ or $i$, $w(z_0) = 0$.

\subsection {Paraxial solutions of rotational symmetric optical system}

Our work focusing mainly on Transmission Electron Microscopes (TEM) and Focused Ion Beam (FIB) instruments, all paraxial solutions will be limited to the action of rotational symmetric lenses. The paraxial contributions of quadrupole lenses through their electrostatic and magnetic second order components $\Phi_2(z)$ or $\Psi_2(z)$ in equation \eqref{eq:equation paraxial} - practically used in TEM and FIB as stigmator components - will be considered as a perturbations of round lenses paraxial solutions. These contributions will correspond to what is known in the literature as axial astigmatism, which shouldn't be confuse with the standard Seidel astigmatism \cite{welford_aberrations_1974}. Then, only  $\Psi_0$ and $\Phi_0$ will be considered in \eqref{eq:equation paraxial}. The paraxial equation that will be considered in this work thus becomes: 

\begin{equation}\label{eq:paraxial equation for round lens}
    \begin{split}
        w''+\frac{\gamma_0}{2\Phi_0^*}(\Phi_0'&+iv_0\Psi_0')w'\\
        &+\frac{\gamma_0}{4\Phi_0^*}\left(\Phi_0''+iv_0\Psi''_0 \right)w = 0
    \end{split}
\end{equation}

\noindent Let's re-define the complex variables using an amplitude/phase representation: 
\begin{eqnarray}
        \chi &=& -\sqrt{\frac{e}{8m_e}}\int_{z_0}^z\frac{\Psi'_0}{\sqrt{\Phi_0^*}}dz\\
        w(z)&=&u(z)e^{i\chi}
\end{eqnarray}

\noindent where $\chi $ corresponds to the Larmor rotation of the electrons \cite{rose_geometrical_2009}. We can then work with only one paraxial equation associated to the evolution of the amplitude $u$: 
\begin{equation}
        u''+\frac{\gamma_0\Phi_0'}{2\Phi_0^*}u'+\frac{\gamma_0}{4\Phi_0^*}\left(\Phi_0''+\frac{e}{2m_e}\Psi'_0{}^2 \right)u = 0
\end{equation}

\noindent The principal $\pi$ and marginal $\alpha$ fundamental solutions of this equation can be calculated using the following boundary conditions:
\begin{equation}\label{eq:definition principal and axial ray}
    \begin{gathered}
        u_\alpha/ u_\alpha(z_0) = 0,u_\alpha'(z_0) = 1\\
        u_\pi/ u_\pi(z_0) = 1,u_\pi'(z_0) = 0
    \end{gathered}
\end{equation}

\noindent Eventually, thanks to the initial conditions $w_0 = x_0+iy_0$ and $w_0' = x_0'+iy_0'$ of any skew ray, the general solution can be written as a simple linear combination:
\begin{equation}
    u = w'_0u_\alpha+w_0u_\pi
\end{equation}

\subsection{Aberrations}\label{ssec:aberration}

All the terms $n>2$ encountered in the polynomial expansion of $\mu$ correspond to deviations added to the paraxial solution of a rotational symmetric optical system. Nevertheless, in such a rotationally symmetric system, the odd $n$ terms in the $\mu^{n}$ expansion disappear. Thus, the first terms to be considered after the paraxial contributions are the 4th-order terms (i.e. $n=4$) - known as primary aberrations or Seidel aberrations. In the following, the term \emph{aberrations} will always refer to wavefront aberrations directly given by the eikonals, while, to avoid any confusion, we will always add the term \emph{transverse} to refer to transverse aberrations.\\

$\mu^{(4)}$ can be written also as a polynomial sum containing five terms known as: spherical, coma, field astigmatism, Petzval curvature and distortion primary aberrations. The expression of each polynomial coefficients can be obtained using the perturbation method originally developed by Seidel for conventional optics and then extended by Glaser and Sturrock, among others, to charged particle optics \cite{welford_aberrations_1974,sturrock_perturbation_1951}. The perturbation method required the used of the eikonal polynomial expansions terms $L^{(n)}$ extracted from equation \eqref{eq: p} using the $\mu^{(n)}$ terms. The Seidel terms coming from the 4th-order contribution of the eikonal correspond to intrinsic aberrations which appear despite the "real" quality of the lens field. Indeed, they are direct consequences of the non-linear behaviour of geometrical optics laws. However, in reality, we will have to additionally consider the parasitic aberrations coming from lenses imperfections or misalignments.\\

Our reference will be always the paraxial solutions considered between the object and Gaussian image planes. The defocus action of a lens can then be defined using a specific second order eikonal term noted $L_d^{(2)}$. The latter corresponds to a difference of second order optical lengths, specifically between the paraxial ray focus at the Gaussian plane and the one considered in a chosen defocused plane. This eikonal linked to the lens defocus is given by \cite{rose_geometrical_2009}: 
\begin{equation}
    L_d^{(2)} = Re\left(\frac{1}{2}C_{df}|w_0'|^2 \right)
\end{equation}

\noindent where:
\begin{equation}
\begin{split}
     C_{df} = \int_{z_0}^z\Bigg[&\sqrt{\frac{\Phi_0^*}{\Phi_0^*(z_0)}} \Big(u_\alpha'^2+\chi'^2u_\alpha^2\\
     &-\frac{\gamma_0}{4\Phi_0^*}\Phi_0''u_\alpha^2\Big) +\frac{e\chi'}{q_0}\Psi_0'u_\alpha^2\Bigg]dz
\end{split}
\end{equation}

\noindent is known as the \emph{defocus}. This expression is correct only for perfect round lenses. However, due to parasitic contributions, some dipolar $\Psi_1^p$, quadrupolar $\Psi_2^p$, or even hexapolar fields $\Psi_3^p$ need be considered. Here, the superscript "$p$" stands for \emph{parasitic}. In practice, to deal with these parasitic contributions, specific electrons optics components such as deflectors, first order and second order stigmators, are inserted along the electron paths. Inside a TEM, magnetic optical elements are added to apply dipolar $\Psi_1$, quadrupolar $\Psi_2$, or hexapolar $\Psi_3$ fields aiming at compensating their parasitic counterparts respectively $\Psi_1^p, \Psi_2^p, \Psi_3^p$. In FIB instruments, electrostatic elements are employed. 

In the following, we will consider $L_p$ to be the eikonal contribution of these parasitic fields and treat it as a perturbation of the rotational symmetric optics eikonal $L^{(2)}$. This means that, in practice, we will always compute the paraxial solutions of \eqref{eq:definition principal and axial ray} instead of using the more general expression given by \eqref{eq:general paraxial fundamental rays}. The multipolar parasitic fields treated as a perturbations of the paraxial solutions will then lead to third order aberrations even in rotational symmetric system - which are added to the fourth ones coming from the standard Seidel intrinsic contributions. Within the framework of transverse aberrations - extracted after differentiating the eikonal function - these contributions will correspond to second order parasitic and third order Seidel aberrations \cite{welford_aberrations_1974}. 

In the next sections, we will briefly dive into this formalism using the nomenclature extracted from \cite{uno_aberration_2005}.

\subsubsection{First, second and third order parasitic aberrations}\label{sssec:parasitic aberrations}

First and second order parasitic aberrations can be expressed using second order variational function given by equations \eqref{eq:mu_m^n parax 0}, \eqref{eq:mu_m^n parax 1}, \eqref{eq:mu_m^n parax 2}, \eqref{eq:mu_e parax 0}, \eqref{eq:mu_e parax 1} and \eqref{eq:mu_e parax 2}. Following the nomenclature of \cite{uno_aberration_2005} and after lengthy mathematical developments - which can be found in detail in \cite{rose_geometrical_2009} - one can define the first and second order eikonal coming from dipole and quadrupole fields as:
\begin{eqnarray}
            L^{(1)}_p+L^{(2)}_p &=& Re\left(A_0\overline{w}_0'+\frac{1}{2}A_1\overline{w}_0'^2 \right) \notag \\
            A_0 &=& \frac{ie}{q_0}\int_{z_0}^zu_\alpha\Psi_1e^{-i\chi}dz \label{Astigmatismeintegralequation}\\
            A_1 &=& \frac{2ie}{q_0}\int_{z_0}^z\Psi_2u_\alpha^2e^{-2i\chi}dz \notag
\end{eqnarray}

\noindent The beam deflection $A_0$ - originating from the dipolar parasitic field $\Psi_1$ - leads to a tilt of the optical axis. The second parasitic aberration $A_1$ is called the axial astigmatism (or first order astigmatism) and induces a difference in paraxial focus between the sagittal and meridional directions. Moreover, the orientation of these sagittal and meridional axis will depend on the phase of $A_1$, which itself depend on the phase of $\Psi_2$ and $\chi$. Third order aberrations can be extracted using the same procedure based on the development of the variational function $\mu$ and lead to the following expression of the third order eikonal generated by dipolar $\Psi_1$ and hexapolar $\Psi_3$ fields: 
\begin{align}
        L^{(3)}_p &= Re\left(\overline{B}_2|w_0'|^2\overline{w}_0'+\frac{1}{3}A_2\overline{w}_0'^3 \right) \label{eq:third order aberrations}\\
        \overline{B}_2 &= -\frac{ie}{4q_0}\int_{z_0}^zdz\Bigg[ \frac{1}{2}\Psi_1''u_\alpha^3+\Psi_1\Big(u_\alpha^2u_\alpha'+i\chi'u_\alpha^3\Big)\Bigg]e^{-i\chi} \notag\\
        A_2 &= \frac{3ie}{q_2}\int_{z_0}^z\Psi_3u_\alpha^3e^{-3i\chi}dz \notag  
\end{align}

\noindent Here, the $B_2$ coefficient is known as the axial coma and is influenced by tilt misalignment with respect to the optical axis. This tilt will be induced by the presence of a dipolar field $\Psi_1$. The variation in second-order focusing along the three directions that are angularly equally distributed, and located perpendicularly to the optical axis, is physically caused by the $A_2$ coefficient associated with second-order astigmatism. This second-order astigmatism is generated by the presence of hexapolar component $\Psi_3$ 

\subsubsection{Fourth order Seidel aberrations}

Eventually, regarding fourth order contribution to the eikonal, we have to separate the Seidel intrinsic contributions - given by the term $L^{(4)}$ - and the one arising from parasitic contributions included in the term $L^{(4)}_p$. We will restrict our analysis to spherical aberration out of the five Seidel aberrations. This primary aberration is associated to the polynomial term depending to the fourth power of the rays angular coordinate $w_0'$. These, fourth order parasitic terms arises from the contribution of quadrupolar fields $\Psi_2$. They lead to various kind of effects that has been extensively studied in the literature \cite{hawkes_springer_1966}. Concerning our study, we will limit the parasitic contribution of quadrupolar field to the star aberration. Then, after heavy mathematical developments detailed in \cite{rose_geometrical_2009}, we obtain the following fourth order eikonal expression:
\begin{equation}
     L^{(4)}+L^{(4)}_p = Re\left(\frac{1}{4}C_s|w_0'|^4+\overline{S}_3|w_0'|^2\overline{w}_0'^2 \right)
\end{equation}

\noindent where the spherical aberration coefficient is defined as:
\begin{equation}\label{csintegralequation}
\begin{split}
    C_s &= \frac{1}{2}\int_{z_0}^zdz\Bigg[\sqrt{\frac{\Phi_0^*}{\Phi_0^*(z_0)}}\Bigg(-\Big(u_\alpha'^2+\chi'^2u_\alpha^2\Big)^2\\
    &-\frac{\gamma_0}{2\Phi_0^*}\Phi_0''u_\alpha^2\Big(u_\alpha'^2+\chi'^2u_\alpha^2\Big) +\frac{1}{16}\Big(\gamma_0\Phi_0''''-\frac{\Phi_0''^2}{\Phi_0^*}\Big)u_\alpha^4 \Bigg)\\
    &-\frac{e}{2q_0}\Psi_0'''\chi'u_\alpha^4 \Bigg]
\end{split}
\end{equation}

\noindent while:
\begin{equation}
     \overline{S}_3 = -\frac{ie}{q_0}\int_{z_0}^zdz\left[\frac{\Psi_2''}{6}u_\alpha^4+\frac{\Psi_2'}{3}\left( u_\alpha'+i\chi'u_\alpha\right)u_\alpha^3\right]e^{-2i\chi} 
\end{equation}

\noindent corresponds to the star aberration. Finally, combining all previous eikonal expressions from first to fourth orders, we obtain the total eikonal expansion:
\begin{equation}\label{eq:L aberration contribution}
    \begin{split}
        L = \Re\Bigg[& A_0\overline{w}_0'+\frac{1}{2}C_{df}|w_0'|^2+\frac{1}{2}A_1\overline{w}_0'^2+\overline{B}_2|w_0'|^2\overline{w}_0'\\
        &+\frac{1}{3}A_2\overline{w}_0'^3+\frac{1}{4}C_s|w_0'|^4+\overline{S}_3|w_0'|^2\overline{w}_0'^2 \Bigg]
    \end{split}
\end{equation}

\noindent In the following, we will show how this equation can be serve to define the generating function which will then be used to apply the catastrophe theory to CPO.

\subsection{Computational methods}\label{compmeth}

\begin{figure}
\centering
\includegraphics[width=\columnwidth]{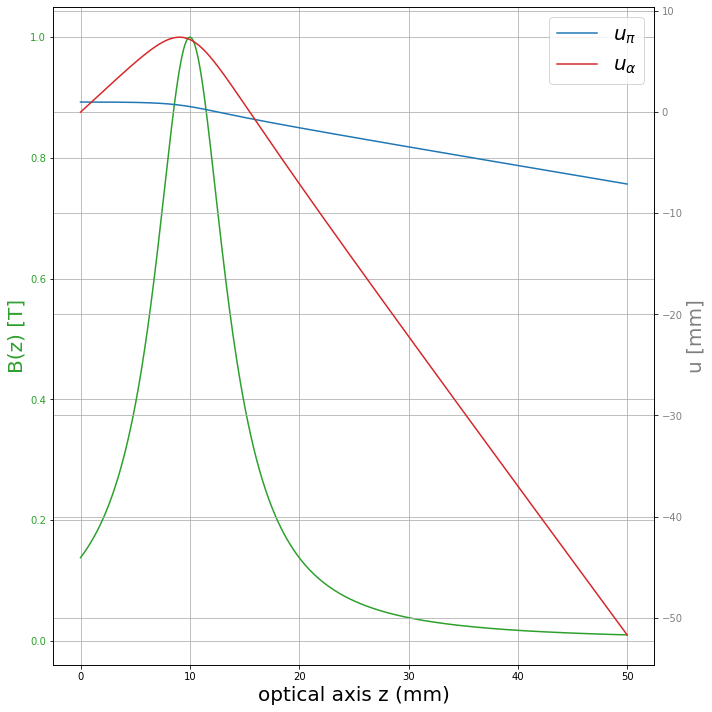}
\caption{\label{fig:RK_glaser}  Paraxial solutions of a magnetic round lens computed using a Runge-Kutta algorithm. A Glaser analytical axial field has been used (in green). $u_{\pi},u_{\alpha}$ are the two fundamentals paraxial solutions. The optical axis is defined as usual by the condition $u=0$.}
\end{figure}

Numerical computations are required to obtain the paraxial solutions and aberration coefficients. We have performed simulations to analyse the data we had obtained specifically using the electron microscope, even though the theory presented above can be used equally to ion and electron optics instruments. Although caustics in ion microscopy is a topic currently being studied, for the sake of this work we will limit our attention to electron microscopy and hence only compute magnetostatic potentials inside the optical system as input datas. Hence, following these calculations, the potentials have to be developed and differentiated along the optical axis in order to determine all the multipolar contributions using the expressions \eqref{eq:phi expension 1}, \eqref{eq:phi expension 2}, \eqref{eq:A expension 1} and \eqref{eq:A expension 2}. Usually, in that situation we first have to solve the Laplace or Poisson equations inside all optical elements using well established numerical methods such as finite difference (FDM), finite element (FEM), or boundary element (BEM) methods (see \cite{hawkes_principles_2018}).This computational methodology would have to be applied if we wished, for example, to extract quantitative data from an optic with a specific geometrical configuration. However, the aim of this work is primarily conceptual, since we simply wish to provide an explanation of the complex geometrical shapes observed within the TEM's caustics, using catastrophe theory. It is therefore essentially a work on concepts and this is why we have chosen to work with simplified analytical forms of the axial fields. In the case of rotationally symmetric lenses, we have chosen to use the Glaser's representation for the axial magnetic flux density given by \cite{hawkes_magnetic_1982}:
\begin{equation}
    B(z) = \frac{B_0}{1+\left(\frac{z-z_l}{a}\right)^2}
\end{equation}
where $B(z) = B_z(x=0,y=0,z) = -\Psi_0'$ correspond to the axial magnetic flux density and $z_l$ is the central position of the lens on the optical axis. In that representation $B_0$ is the maximum strength of the axial magnetic flux density  and $a$ is the half peak width of this so-called Glaser's bell-shaped distribution. Furthermore we will consider the multipolar fields $\Phi_n$ and $\Psi_n$ components constants along the optical axis. From a physical standpoint, this implies that we will consistently ignore the impact of fringing fields. We should avoid such an approximation if we need to extract quantitative data from an optical setup that contains a strong multipolar field. However, in our work, we will always consider these fields as only a correction from the rotationally symmetric ones, and conceptually speaking, this approximation will not alter any of the study's qualitative conclusions.

Finally to solve the paraxial equation (\ref{eq:paraxial equation for round lens}) we used an home-made \texttt{Python} code based on Runge-Kutta numerical method (see for instance this introductory book \cite{appel_mathematics_2007}).
Figure (\ref{fig:RK_glaser}) reports the two paraxial fundamental rays computed by this software using the bell shape Glaser's field as an input field for the equation. This axial field shape has also been represented by the green curve computed using the parameters $z_l = 10$mm, $a = 4$mm, $B_0 = 1$T.

\section{Catastrophe charged-particle optics}\label{catastropheCPO}

The contribution of M. Berry and J. Nye to catastrophe optics - presented in the introduction - will serve as the guideline to extract the generating function of our CPO problem. We start our development with \eqref{eq:straght line CCO} established that in the case of light. As a reminder, we used it to calculate the optical path length between the exit pupil plane, where the wavefront is described, and the caustic observation plane, see Fig.~\ref{fig2}(b). In the case of CCPO, instead of using $l$ to build the generating function, we will simply use the normalized point eikonal $L$ as defined by the \eqref{eq:variationnal function mu}. The optical path found in conventional optics is nevertheless equivalent to this function. The expansion of this function, given by \eqref{eq:L aberration contribution} will serve to derive a canonical form of the generating function as prescribed by the catastrophe theory.\\

\subsection{Charged-particles optical setup}

First, it is important to provide an overview of the CPO setup used for our studies. Again, we stress that our analysis will only be performed using the data acquired on the TEM, while the one acquired on the FIB will only be provided as illustration and leave the corresponding details for a future article.

\begin{figure}\
\centering
\includegraphics[width=0.5\textwidth]{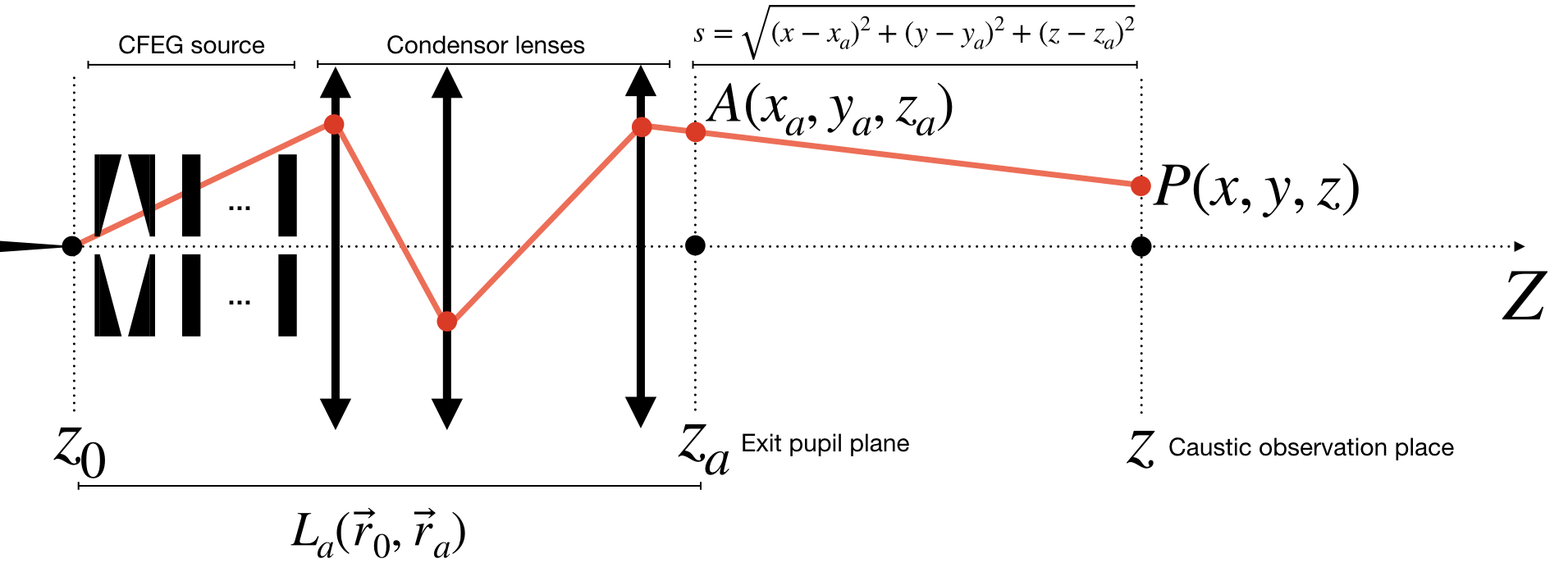}
\caption{\label{fig5} Electron optical path between the exit pupil plane of the I2TEM illumination system and the caustic observation plane}
\end{figure}

The caustics were acquired on the Hitachi I2TEM microscope of CEMES-Toulouse (In Situ Interferometry Transmission Electron Microscope) \cite{houdellier_coherent_2023}. This instrument is based on an HF3300 column of the Japanese manufacturer Hitachi equipped with a $300$ keV cold field emission source located before a three condensor lenses system, a double stage objective lens and an imaging aplanatic corrector designed and manufactured by the german company CEOS gmbH \cite{muller_aplanatic_2011}. The caustics have been recorded using a CMOS-based detector manufactured by the US company Gatan inc. The schematic overview of the probe forming system, which contains the only components of the larger instrument that will be taken into account for our generating function calculation, is reported in Fig.~\ref{fig5}. 

In the above-described instrument, the optical path will be simply determined by taking into account all the magnetostatic potentials $\Psi$ encountered in the probe forming lens system. In the FIB scenario however, the optical system's electrostatic potential $\Phi$ would have to be considered instead. Apart from that, the method described in the next section can be applied exactly as it is. 

\subsection{CCPO formalism}

\subsubsection{The generating function}

We shall divide the optical path into two separate functions using a geometrical method, which will prove as convenient for defining the generating function. This approach was first presented by H. Rose in \cite{rose_geometrical_2009}.

The two geometrical areas are reported on Fig.~\ref{fig5} and will be addressed separately. There, $z$ and $z_0$ indicate the locations of the caustic observation plane and the electron source, respectively, along the optical axis. The exit pupil plane's coordinate of the probe forming optics is noted $z_a$. The point $A(x_a,y_a,z_a)$ in the exit pupil plane and the point $Q(x,y,z)$ in the caustic observation plane are connected by a random electron trajectory. In order to simplify our theoretical strategy, we shall consistently consider the exit pupil plane as being outside of any electromagnetic field generated by the optical system. 

Combining $L_a$ and $s$ - defined on Fig.~\ref{fig5} - enables us to determine the optical path $L$ as $L=L_a+s$. In this expression, $L_a$ corresponds to the optical path between $z_0$ and the exit pupil plane $z_a$, while $s$ corresponds to the one located between the exit pupil plane $z_a$ and $z$ - which is considered in the field-free zone. The calculation of $L_a$ is then done by integrating the variational function $\mu$ between $z_0$ and $z_a$ and by considering the axial magnetic potential and their derivatives along the optical axis \eqref{eq:variationnal function mu}. The wavefront aberration function is also included in this first function. The second component of the optical path - $s$ - is expressed by taking into account a straight line connecting $A$ and $P$. In order to eliminate the square root appearing in the expression, we apply the far-field approximation as done for conventional catastrophe optics in the introduction:
\begin{eqnarray}
        s &=& \sqrt{(x-x_a)^2+(y-y_a)^2+(z-z_a)^2}\\
         &\approx& z-z_a+\frac{(x-x_a)^2}{2z}+\frac{(y-y_a)^2}{2z}+\frac{z_a^2}{2z}
\end{eqnarray}

The optical path $L_a$ will be defined by the equation (\ref{eq:L aberration contribution}). This function is written as a sum of polynomial functions of $w_0'$. However, similar to conventional optics, we have to choose the pupil plane coordinates $(x_a,y_a)$ as the states variables for our generating function. In order to express the optical path $L_a$ using the complex pupil plane coordinates $w_a = x_a+iy_a$ we will have first to rewrite it after converting these variables. We have first to set $u_a = u(z_a)$, $u_{\alpha a} = u_\alpha(z_a)$ and $\chi_a = \chi(z_a)$. Given that, by definition, the paraxial principal trajectory crosses the optical axis at the centre of the exit pupil plane, with $u_\pi(z_a) = 0$, one has:
\begin{eqnarray}
        u_a &=& w_0'u_{\alpha a}\notag\\
        w_a &=& u_{\alpha a}e^{i\chi_a} \label{eq:expr pupil plane coordinates}\\
        \implies w_0' &=& \frac{w_a}{u_{\alpha a}}e^{-i\chi_a}\notag
\end{eqnarray}

\noindent Equation \eqref{eq:L aberration contribution} can now be used to express the optical path $L_a$ as the polynomial sum of $w_a$ and its complex conjugate by substituting $w_0'$ by it's expression \eqref{eq:expr pupil plane coordinates}. Specifically, we obtain: 
\begin{equation}
\begin{split}
        L_a &= \Re\Bigg[\frac{A_0}{u_{\alpha a}}e^{i\chi_a}\overline{w}_a+\frac{1}{2}\frac{C_{df}}{u_{\alpha a}^2}|w_a|^2+\frac{1}{2}\frac{A_1}{u_{\alpha a}^2}e^{2i\chi_a}\overline{w}_a^2\\
        &+\frac{\overline{B}_2}{u_{\alpha a}^3}e^{i\chi_a}|w_a|^2\overline{w}_a+\frac{1}{3}\frac{A_2}{u_{\alpha a}^3}e^{3i\chi_a}\overline{w}_a^3+\frac{1}{4}\frac{C_s}{u_{\alpha a}^4}|w_a|^4\\
        &+\frac{\overline{S}_3}{u_{\alpha a}^4}e^{2i\chi_a}|w_a|^2\overline{w}_a^2 \Bigg]
\end{split}
\end{equation}

\noindent We now expand this expression in order to obtain a $x_a,y_a$ polynomial expression. For clarity, we will proceed with the expansion order per order:
\begin{equation}
    L_a=L_a^{(1)}+L_a^{(2)}+L_a^{(3)}+L_a^{(4)}
\end{equation}

\noindent where the first order term can be expanded as:
\begin{eqnarray}
     L_a^{(1)} &=& \Re\left[ \frac{A_0}{u_{\alpha a}}e^{i\chi_a}\overline{w}_a\right]\\
     &=& \Re\left[ \frac{A_0}{u_{\alpha a}}e^{i\chi_a}\right]x_a+\Im\left[ \frac{A_0}{u_{\alpha a}}e^{i\chi_a}\right]y_a
\end{eqnarray}

\noindent The second order term is developed as:
\begin{align}
      L_a^{(2)} &= \Re\left[\frac{1}{2}\frac{C_{df}}{u_{\alpha a}^2}|w_a|^2+\frac{1}{2}\frac{A_1}{u_{\alpha a}^2}e^{2i\chi_a}\overline{w}_a^2\right]\\
      \begin{split}
        &= \frac{1}{2}\frac{C_{df}}{u_{\alpha a}^2}(x_a^2+y_a^2)+\Re\left[ \frac{1}{2}\frac{A_1}{u_{\alpha a}^2}e^{2i\chi_a}\right](x_a^2-y_a^2)\\
        &+\Im\left[ \frac{A_1}{u_{\alpha a}^2}e^{2i\chi_a}\right]x_ay_a
        \end{split}
\end{align}

\noindent The third order term reads:
\begin{align}
        L_a^{(3)} &= \Re\left[\frac{\overline{B}_2}{u_{\alpha a}^3}e^{i\chi_a}|w_a|^2\overline{w}_a+\frac{1}{3}\frac{A_2}{u_{\alpha a}^3}e^{3i\chi_a}\overline{w}_a^3 \right]\\
        \begin{split}
        &=\Re\left[\frac{\overline{B}_2}{u_{\alpha a}^3}e^{i\chi_a}+\frac{1}{3}\frac{A_2}{u_{\alpha a}^3}e^{3i\chi_a}\right]x_a^3\\
        &+\Im\left[\frac{A_2}{u_{\alpha a}^3}e^{3i\chi_a}-\frac{\overline{B}_2}{u_{\alpha a}^3}e^{i\chi_a}\right]x_a^2y_a\\
        &+\Re\left[\frac{\overline{B}_2}{u_{\alpha a}^3}e^{i\chi_a}-\frac{A_2}{u_{\alpha a}^3}e^{3i\chi_a}\right]x_ay_a^2\\
        &-\Im\left[\frac{\overline{B}_2}{u_{\alpha a}^3}e^{i\chi_a}+\frac{1}{3}\frac{A_2}{u_{\alpha a}^3}e^{3i\chi_a}\right]y_a^3
        \end{split}
\end{align}

\noindent Finally, the fourth order term can be expanded as:
\begin{align}
    L_a^{(4)} &= \Re\left[ \frac{1}{4}\frac{C_s}{u_{\alpha a}^4}|w_a|^4+\frac{\overline{S}_3}{u_{\alpha a}^4}e^{2i\chi_a}|w_a|^2\overline{w}_a^2\right]\\
    \begin{split}
    &= \left( \frac{1}{4}\frac{C_s}{u_{\alpha a}^4}+\Re\left[\frac{\overline{S}_3}{u_{\alpha a}^4}e^{2i\chi_a} \right]\right)x_a^4\\
    &-2\Im\left[\frac{\overline{S}_3}{u_{\alpha a}^4}e^{2i\chi_a} \right]x_a^3y_a\\
    &+ \frac{1}{2}\frac{C_s}{u_{\alpha a}^4}x_a^2y_a^2-2\Im\left[\frac{\overline{S}_3}{u_{\alpha a}^4}e^{2i\chi_a} \right]x_ay_a^3\\
    &+\left( \frac{1}{4}\frac{C_s}{u_{\alpha a}^4}-\Re\left[\frac{\overline{S}_3}{u_{\alpha a}^4}e^{2i\chi_a} \right]\right)y_a^4
    \end{split}
\end{align}

\noindent Following the formalism of J. Nye \cite{nye_natural_1999,nye_relation_2005}, the Fermat's potential $\Phi_f$ - which will be the ground on which the generating function will be built - can now be expressed using the formula : 
\begin{equation}\label{generatingfunctionCPO}
   \Phi_f = L_a+\frac{x_a^2+y_a^2}{2z}-\frac{xx_a+yy_a}{z}
\end{equation}

\noindent Finally, in order to write down the generating function, it's necessary to display all the polynomials define by the different powers of the variables $(x_a,y_a)$. Gathering all the orders previously developed in one expression, we find the Fermat's potential $\Phi_f$ :
\begin{widetext}
\begin{equation}\label{fermatspotential}
\begin{split}
   \Phi_f  &= \left(\Re\left[ \frac{A_0}{u_{\alpha a}}e^{i\chi_a}\right]-\frac{x}{z}\right)x_a+\left(\Im\left[ \frac{A_0}{u_{\alpha a}}e^{i\chi_a}\right]-\frac{y}{z}\right)y_a+\left(\frac{1}{2}\frac{C_{df}}{u_{\alpha a}^2}+\frac{1}{2z}\right)(x_a^2+y_a^2)\\
   &+\Re\left[ \frac{1}{2}\frac{A_1}{u_{\alpha a}^2}e^{2i\chi_a}\right](x_a^2-y_a^2)+\Im\left[ \frac{A_1}{u_{\alpha a}^2}e^{2i\chi_a}\right]x_ay_a+\Re\left[\frac{\overline{B}_2}{u_{\alpha a}^3}e^{i\chi_a}+\frac{1}{3}\frac{A_2}{u_{\alpha a}^3}e^{3i\chi_a}\right]x_a^3\\
   &+\Im\left[\frac{A_2}{u_{\alpha a}^3}e^{3i\chi_a}-\frac{\overline{B}_2}{u_{\alpha a}^3}e^{i\chi_a}\right]x_a^2y_a+\Re\left[\frac{\overline{B}_2}{u_{\alpha a}^3}e^{i\chi_a}-\frac{A_2}{u_{\alpha a}^3}e^{3i\chi_a}\right]x_ay_a^2-\Im\left[\frac{\overline{B}_2}{u_{\alpha a}^3}e^{i\chi_a}+\frac{1}{3}\frac{A_2}{u_{\alpha a}^3}e^{3i\chi_a}\right]y_a^3\\
        &+\left( \frac{1}{4}\frac{C_s}{u_{\alpha a}^4}+\Re\left[\frac{\overline{S}_3}{u_{\alpha a}^4}e^{2i\chi_a} \right]\right)x_a^4-2\Im\left[\frac{\overline{S}_3}{u_{\alpha a}^4}e^{2i\chi_a} \right]x_a^3y_a+ \frac{1}{2}\frac{C_s}{u_{\alpha a}^4}x_a^2y_a^2-2\Im\left[\frac{\overline{S}_3}{u_{\alpha a}^4}e^{2i\chi_a} \right]x_ay_a^3\\
        &+\left( \frac{1}{4}\frac{C_s}{u_{\alpha a}^4}-\Re\left[\frac{\overline{S}_3}{u_{\alpha a}^4}e^{2i\chi_a} \right]\right)y_a^4
\end{split}
\end{equation}
\end{widetext}

\noindent By comparing this expression to the canonical form of the seven catastrophes, one can understand that all aberration coefficients will have to be treated as control variables. However, in this first work, we have chosen to restrict our analysis to the codimension that has already been covered by R. Thom's and V. Arnold's works and, as a consequence we have decided to discard the star aberration $S_3$ from \eqref{fermatspotential}. Indeed, if one includes this aberration the generating function becomes excessively complex. Nevertheless, given that our CPO system is mainly limited by the spherical aberration of the condensor lenses, this approximation remains reasonable. To fit our expression within the canonical catastrophes (see table \ref{tab:elementary catastrophes}) - which does not exhibits coefficients in the germ polynomial expressions - the Fermat's potential \eqref{fermatspotential} has to be divided by the highest order coefficient of the expansion, which corresponds to the spherical aberration. Eventually, we obtain the final expression for the generating function $\Phi$:
\begin{equation}
    \begin{split}
   \Phi&(x_a,y_a,C_1,C_2,C_3,C_4,C_5,C_6,C_7,C_8,C_9) =\\
   &C_1x_a+C_2y_a+C_3(x_a^2+y_a^2)+C_4(x_a^2-y_a^2)+C_5x_ay_a\\
        &+C_6x_a^3+C_7x_a^2y_a+C_8x_ay_a^2+C_9y_a^3+x_a^4+2x_a^2y_a^2+y_a^4
    \end{split}
\end{equation}

\noindent in which the following control variables have been defined:
\begin{equation}\label{eq:control variables annexes 1}
C_1 = 4\left(\Re\left[ u_{\alpha a}^3\frac{A_0}{C_s}e^{i\chi_a}\right]-\frac{x}{zC_s}u_{\alpha a}^4\right)
\end{equation}

\begin{equation} \label{eq:control variables annexes 2}
C_2 = 4\left(\Im\left[ u_{\alpha a}^3\frac{A_0}{C_s}e^{i\chi_a}\right]-\frac{y}{zC_s}u_{\alpha a}^4\right)
\end{equation}

\begin{equation}
C_3 = 2\left(u_{\alpha a}^2\frac{C_{df}}{C_s}+\frac{u_{\alpha a}^4}{zC_s}\right)
\end{equation}

\begin{equation}  \label{eq:control variables annexes 4}
C_4 = 2\Re\left[u_{\alpha a}^2\frac{A_1}{C_s}e^{2i\chi_a}\right]
\end{equation}

\begin{equation}  \label{eq:control variables annexes 5}
C_5 = 4\Im\left[ u_{\alpha a}^2\frac{A_1}{C_s}e^{2i\chi_a}\right]
\end{equation}

\begin{equation}
C_6 = 4\Re\left[u_{\alpha a}\frac{\overline{B}_2}{C_s}e^{i\chi_a}+\frac{u_{\alpha a}}{3}\frac{A_2}{C_s}e^{3i\chi_a}\right]
\end{equation}

\begin{equation}
C_7 = 4\Im\left[u_{\alpha a}\frac{A_2}{C_s}e^{3i\chi_a}-u_{\alpha a}\frac{\overline{B}_2}{C_s}e^{i\chi_a}\right]
\end{equation}

\begin{equation}
C_8 = 4\Re\left[u_{\alpha a}\frac{\overline{B}_2}{C_s}e^{i\chi_a}-u_{\alpha a}\frac{A_2}{C_s}e^{3i\chi_a}\right]
\end{equation}

\begin{equation} \label{eq:control variables annexes 9}
C_9 = -4\Im\left[u_{\alpha a}\frac{\overline{B}_2}{C_s}e^{i\chi_a}+\frac{u_{\alpha a}}{3}\frac{A_2}{C_s}e^{3i\chi_a}\right]
\end{equation}

\noindent Consequently, even when considering the simplest scenario conventionally encountered during TEM experiments, the codimension of our generating function is 9 - a scenario beyond the seven elementary catastrophes studied by Thom's catastrophe theory. We have therefore to consider the extension of the catastrophe theory developed by V. Arnold \cite{arnold_catastrophe_1992}. Specifically, the germ of our generating function is:
\begin{equation}
    g(x_a,y_a) = x_a^4+2x_a^2y_a^2+y_a^4
\end{equation}

\noindent Thus, our generating function corresponds to the so-called $X^9$ catastrophe of Arnold's theory, which is characterized by the germ $g =x_a^4+Kx_a^2y_a^2+y_a^4$ with $K=2$ \cite{callahan_special_1982,arnold_critical_1975}.\\ 

\begin{figure}
\centering
\includegraphics[width=\columnwidth]{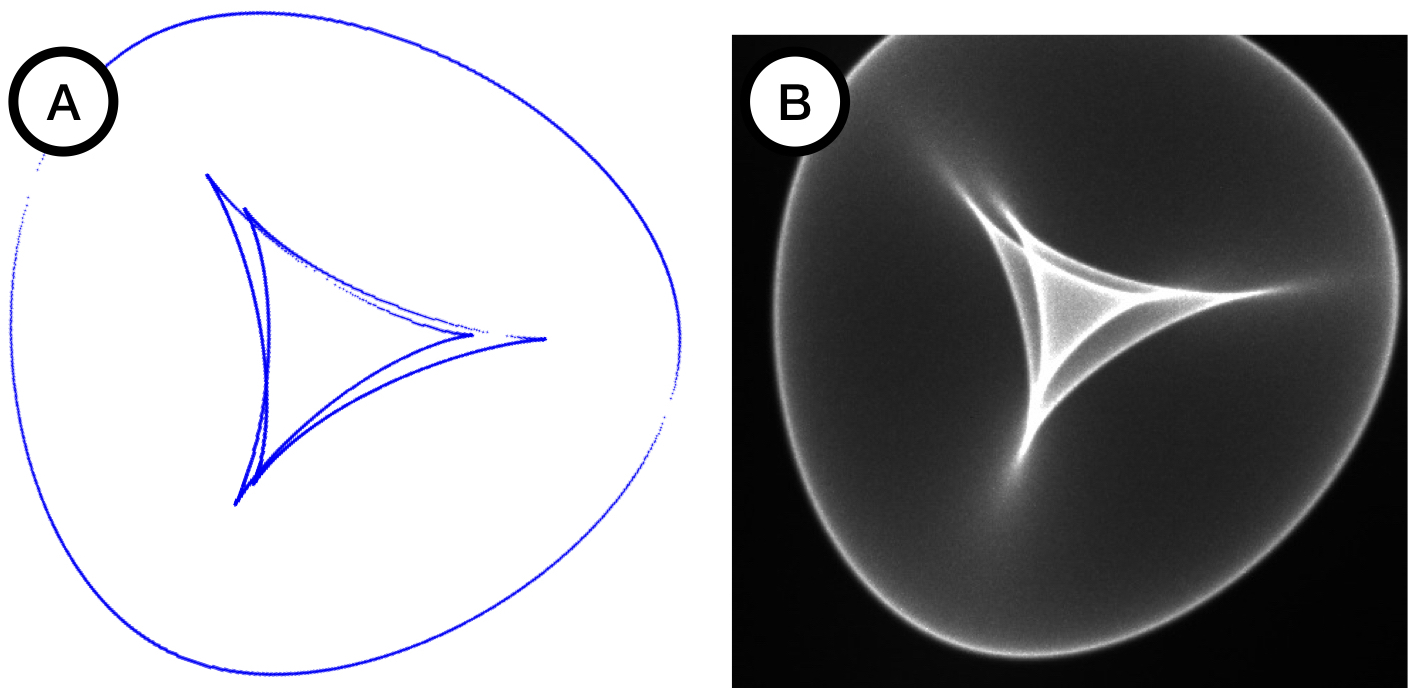}
\caption{\label{fig6} Left : Simulated caustics projected using a $X^9$ generating function (inspired by Nye's pioneering work \cite{nye_natural_1999}). Right : Caustics taken using the I2TEM instrument.}
\end{figure}

The $X^9$ catastrophe can have various shapes and geometry depending on the values of the control variables. Nevertheless, based on our observations, the $X^9$ catastrophe seem to fit all measured TEM caustics topologies, therefore we have chosen to base all of our investigation on this generating function. To visually illustrate our strategy, we report on Fig. \ref{fig6} a qualitative comparison between one caustic observed in the I2TEM using a strong $A_2$ aberration - applied using short hexapoles located between the condensor lenses - and one of the possible $X^9$ symmetry simulated using our code based on this work and strongly inspired from a situation described in J. Nye's book \cite{nye_natural_1999}. 

\subsubsection{Caustics}

The stationary conditions are obtained by applying (\ref{eq:critical set of Phi CCO}) to $\Phi$ :
\begin{equation}\label{eq:fermat principle CEO}
    \frac{\partial  \Phi}{\partial x_a} = \frac{\partial  \Phi}{\partial y_a} = 0
\end{equation}

\noindent This operation is used to define our system's equilibrium surface, from which any possible trajectories of a charged particle can be derived. In particular, as shown in section \ref{SEC:introduction_catastrophe_theory}, the caustics are then given by the singularities of this surface. The latter can be found by calculating the roots of the Hessian determinant:
\begin{equation}\label{eq:Hessain det CEO}
    \frac{\partial^2\Phi}{\partial x_a^2}\frac{\partial^2\Phi}{\partial y_a^2}-\left( \frac{\partial^2\Phi}{\partial x_a\partial y_a}\right)^2 = 0
\end{equation}

\noindent We shall therefore obtain the observed caustics from the projection of all singularities encountered in the equilibrium surface onto the control space.\\ 

However, equation \eqref{eq:Hessain det CEO} presents an issue since the Hessian's determinant is a fourth-order polynomial function. As a result, in most situations, it will be impossible to obtain analytical solutions for the Hessian determinant's roots.

Numerical computations constitute one possible solution that we will use to solve this issue. The aberration coefficients could indeed be computed from Glaser's analytic model of the axial magnetic field of round lenses, as presented in section \eqref{sec:OPC}. Additional quadrupoles $\Psi_2$ or hexapoles $\Psi_3$ field strengths can also be defined analytically by neglecting all the fringing field effects \cite{wollnik_optics_2014}. The knowledge of these aberrations enable the computation of the control variables using equation \eqref{eq:control variables annexes 1} to \eqref{eq:control variables annexes 9}. The Hessian determinant equation $\text{det}(H)=0$ \eqref{eq:Hessain det CEO} is then solved using an home-made \texttt{Python} code employing several numerical approximations.

Nevertheless, we will show in the next section that - in some specific situations - one can derive exact analytical solutions of $\text{det}(H)=0$. These expressions will enable us to establish relationships between the aberrations and the caustic shape. The latter will reveals complex relations hidden inside the formalism of charged particles optics which would have been impossible to easily extract without the catastrophe theory formalism.

\subsection{Application: Numerical derivation of caustic geometry from optical calculations}\label{sec:application1}

We begin by solving equation \eqref{eq:Hessain det CEO} using a generating function whose coefficients were derived numerically from integral expressions of the aberrations. Estimations have been performed for the optical system depicted on Fig.~\ref{fig5} corresponding to the illumination system of the I2TEM microscope. In addition, the axial field of the condenser lenses are expressed using Glaser's method. With these conditions, derivations can therefore be analytically performed very easily. Thus, procedures described in section \ref{compmeth} are applied to determine the paraxial solutions and the aberrations coefficients. We will describe the results of these calculations in the rest of this section.\\

\begin{figure}
\centering
\includegraphics[width=\columnwidth]{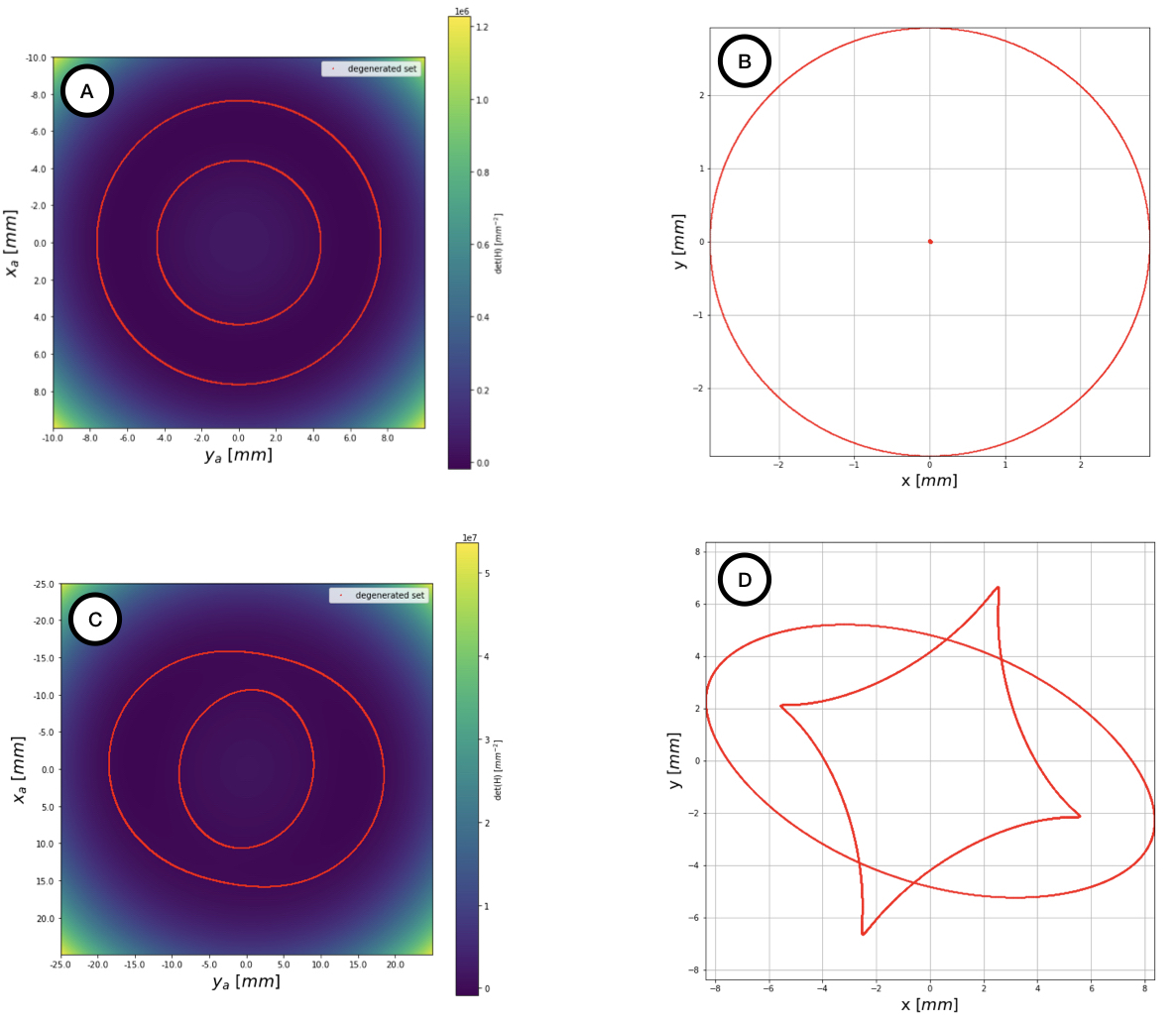}
\caption{\label{fig7}  A : Hessian determinant and B : caustic observed in the observation plane $z$. calculation are realised for a pure rotationally symmetric configuration. C: hessian determinant and D : caustic determined after a quadrupole $\Psi_2$ field component was added.}
\end{figure}

Fig.~\ref{fig6}(a) shows the Hessian determinant mapped as a function of the pupil plane coordinates when considering only rotationally symmetric components of the fields i.e. $\Psi_0\neq 0$ and $\Psi_1=\Psi_2=...=0$. The spherical aberration coefficient $C_s$ is determined with \eqref{csintegralequation}, and we consider no defocus $C_{df}=0$. In that case the generating function contains only the first three control variables $C_1,C_2,C_3$. The Hessian determinant can moreover be calculated with setting $C_1,C_2=0$ in $\Phi$ since these two control variables will nevertheless disappear due to the double derivative in \eqref{eq:Hessain det CEO}. In practice, the zeros of the Hessian determinant - i.e. the degenerated set - are numerically extracted from the sign changes in the function when mapping it on the pupil plane space ($x_a$,$y_a$). These points are marked by the red dots on the $\text{det}(H)=0$ map  (see Fig.~\ref{fig7}(a)). 

One can observe two subsets of these roots mapped along two circles of different radius centred on $(0,0)$. Applying Fermat's principle \eqref{eq:fermat principle CEO} for the rays defined by the two degenerative sets in the pupil plane, we can now map the caustic in the control space $C_1,C_2$, which is equivalent to the $(x,y)$ usual space. To plot the caustic on Fig.~\ref{fig6}(b) we have set $z=z_g$ for the Gaussian plane (i.e. $C_{df}=0$). A typical beam shape - that has a first caustic point at $(0,0)$ and a second caustic that corresponds to a circle of $(0,0)$ centre - is observed. The scale is given in $mm$ since we neglected the effect of the demagnification coming from the objective lens located after the condenser lenses.

In a second step, we add a small quadrupolar field $\Psi_2=0.01$ T/m corresponding to an order of magnitude usually encountered in condenser stigmators. The two-fold astigmatism $A_1$ can then be estimated using the expression \eqref{Astigmatismeintegralequation} by integrating the paraxial trajectory computed over our rotationally symmetric system. The Hessian determinant is then mapped onto the pupil plane space and the degenerated sets are numerically estimated as previously done. Fig.~\ref{fig6}(c) again displays two subsets with a slight elliptical distortion. Going back to the control space $(x,y)$ using the Fermat's principle and setting $z=z_g$, one can observe the same caustic behaviour as on Fig.~\ref{fig1}. Indeed, on Fig.~\ref{fig6}(d), one can clearly observe two curves: the first one with a shape of a standard ellipse and a second one with a much more complex behaviour composed of four cusps tips located along specific orientation of the first ellipse.\\

By implementing this method, we have demonstrated that if the paraxial and aberration parameters of the system are first calculated, it is then possible to determine the caustic shape. Unfortunately, the estimation of optical properties takes a lot longer than the actual caustic computation. Indeed, the catastrophe theory formalism makes caustic extraction quite easy. Therefore, remembering that we began with very simple analytical forms for the axial potentials, the approach remains nevertheless limited by the estimation of the optical properties. In reality, starting with real lenses parameters (e.g. shapes, magnetization state of the pole pieces, coils currents) we will have to use finite element methods to numerically compute these potentials which will drastically increase the overall computation time. The major benefit of catastrophe theory - which lies in the simplicity of its physical approach to the understand of caustic formation - is compromised by this complex numerical process, which begins with the computation of the potentials and their derivatives and concludes with the estimation of the function $L$. Therefore, we do not find the method particularly interesting in its current state, even compared to the raw solution that would require calculating each trajectory separately to extract the geometric location of the caustic.\\ 

Nevertheless, one can implement an approach based only on the properties of the generating function, without any \textit{a priori} knowledge of the optical system parameters. Catastrophe theory actually allows us to treat any optical problem mathematically as an exercise of topology defined by state and control variables, regardless of the details of their reality. Of course, if one wish to perform a link with an experiment through a measurement - for example the size of a caustic - one has to estimate the dimensions of the control variables, and as a consequence, to compute all the optical properties (paraxial and aberrations). We thus fall back into the previous pitfall. On the other hand, one can connect our calculation to observables by comparing dimensionless data extracted from the mathematical treatment with the experiment, for example ratios between the control variables. This method constitutes, in fact, a very effective strategy that we will develop in the next section. In particular, we will see that catastrophe theory offers the possibility of an analytical treatment of optical problems that are far too complex to be analytically tackled with standard CPO formalism. 

\subsection{Application: analytical derivation of caustic geometry using catastrophe theory with no \textit{a priori} knowledge } \label{sec:application2}

\subsubsection{System with rotationally symmetry}
Our starting point will be the rotationally symmetric system that restricts the optical path expansion to the two coefficients $C_s$ and $C_{df}$.
Considering first $C_s=0$ the generating function can be written in its most simplest form as follow :
\begin{equation}
    \Phi (x_a,y_a,C_1,C_2,C_3) = C_1x_a+C_2y_a+C_3(x_a^2+y_a^2)
\end{equation}

\noindent with:
\begin{eqnarray}
    C_1 &=& -\frac{x}{z}\\
    C_2 &=& -\frac{y}{z}\\
    C_3 &=& \frac{C_{df}}{2u_{\alpha a}^2}+\frac{1}{2z} \label{C3control}
\end{eqnarray}

\noindent All control variables have been previously described in equations \eqref{eq:control variables annexes 1} to \eqref{eq:control variables annexes 9}. The equilibrium surface is therefore given by:
\begin{empheq}[left={\empheqlbrace\,}]{align}
        \frac{\partial \Phi}{\partial x_a} &= C_1+2C_3x_a = 0 \label{eq:equilibrium surface no aberrations 1}\\
        \frac{\partial \Phi}{\partial y_a} &= C_2+2C_3y_a = 0 \label{eq:equilibrium surface no aberrations 2}
\end{empheq}

\noindent In that situation, $C_3 = 0$ provides the only root of the Hessian, which - when plugged into \eqref{eq:equilibrium surface no aberrations 1} and \eqref{eq:equilibrium surface no aberrations 2} - yields $(C_1,C_2) = (0,0)$ which correspond to $x=y=0$ in real space. Hence, all rays defined by any set of pupil coordinates $(x_a,y_a)$ will cross at this point. This is the well know definition of paraxial image plane located at $z_g=-u_{\alpha a}^2/C_{df}$. This simple example has the benefit of showing how analytical manipulation of the generating function can yield significant optical knowledge.\\

Adding the spherical aberration, the generating function becomes: 
\begin{equation}
\begin{split}
       \Phi(x_a,y_a,C_1,C_2,C_3) &= C_1x_a+C_2y_a+C_3(x_a^2+y_a^2)\\
       &+x_a^4+2x_a^2y_a^2+y_a^4
\end{split}
\end{equation}

\noindent As done previously, one can calculate the equilibrium surface by differentiating $\Phi$ :
\begin{empheq}[left={\empheqlbrace\,}]{align}
        \frac{\partial \Phi}{\partial x_a} &= C_1+2C_3x_a +4x_a^3+4x_ay_a^2 = 0 \label{eq:equilibrium map spherical aberration 1}\\
        \frac{\partial  \Phi}{\partial y_a} &= C_2+2C_3y_a+4x_a^2y_a+4y_a^3 = 0 \label{eq:equilibrium map spherical aberration 2}
\end{empheq}

\noindent which leads to the degenerate set:
\begin{equation}\label{eq:det(H) just C_s}
    \begin{split}
        \text{det}(H[ \Phi]) = 48x_a^4+& 96x_a^2y_a^2+48y_a^4\\
        &+32C_3(x_a^2+y_a^2)+4C_3^2 = 0
    \end{split}
\end{equation}

\noindent Carrying out a change of variable $X=x_a^2+y_a^2$, equation (\ref{eq:det(H) just C_s}) can be re-written as:
\begin{equation}
    48X^2+32C_3X+4C_3^2 = 0
\end{equation}

\noindent which admits two solutions:
\begin{empheq}[left={\empheqlbrace\,}]{align}
        X_1 &= \frac{-C_3}{2}\\
        X_2 &= \frac{-C_3}{6}
\end{empheq}

\noindent By definition, $X$ has to be positive, thus equation \eqref{eq:det(H) just C_s} does not have any solution if $C_3>0$. Therefore, a caustic can be observed if and only if $C_3\leq0$, for instance before the Gaussian plane $z<z_g$ when $C_s>0$. We then retrieve a well known result but expressed within a more general condition.

One can also identify the location of the two roots mapped into the pupil plane using the equation of a circle $x_a^2+y_a^2 = X_i$ ($i=(1,2)$). We have thus demonstrated analytically the origin of the two red circles first encountered on Fig.~\ref{fig7}(a). The small one is associated to the $X_2$ solution corresponding to a radius $\sqrt{-C_3/6}$, while the larger one corresponds to the $X_1$ solution defined by a radius of $\sqrt{-C_3/2}$. Going back to the equilibrium surface \eqref{eq:equilibrium map spherical aberration 1} and \eqref{eq:equilibrium map spherical aberration 2}, we may determine the caustics generated by these $X_1$ and $X_2$ solutions by projecting their locations in the equilibrium surface onto the control space. To this aim, let's define $(C_{1i},C_{2i})$ the set of control variables which define the caustic function extracted from the solution $X_i$. Knowing that $x_a^3 = x_aX_i-y_a^2x_a$ and $y_a^3 = y_aX_i-x_a^2y_a$, we can write the equilibrium surface as: 
\begin{empheq}[left={\empheqlbrace\,}]{align}
        \frac{\partial \Phi}{\partial x_a} &= 4x_aX_i+2C_3x_a+C_{1i}\\
        \frac{\partial \Phi}{\partial y_a} &= 4y_aX_i+2C_3y_a+C_{2i}
\end{empheq}

\noindent Hence the caustic generated by the $X_1$ solution corresponds to a point located at $(C_1,C_2) = (0,0)$ i.e. $x = y = 0$. In order to extract the caustic generated by the $X_2$ solution, we have to solve the following system:
\begin{empheq}[left={\empheqlbrace\,}]{align}
        \frac{\partial \Phi}{\partial x_a} &= \frac{4}{3}x_aC_3+C_{12}  = 0\\
        \frac{\partial \Phi}{\partial y_a} &=\frac{4}{3}y_aC_3+C_{22} = 0
\end{empheq}

\noindent which gives:
\begin{empheq}[left={\empheqlbrace\,}]{align}
        x_a &= -\frac{3C_{12}}{4C_3}\\
        y_a &= -\frac{3C_{22}}{4C_3}
\end{empheq}

\noindent and therefore:
\begin{equation}\label{causticcircle}
    C_{12}^2+C_{22}^2 = -\frac{8}{27}C_3^3
\end{equation}

\begin{figure}
\centering
\includegraphics[width=0.5\textwidth]{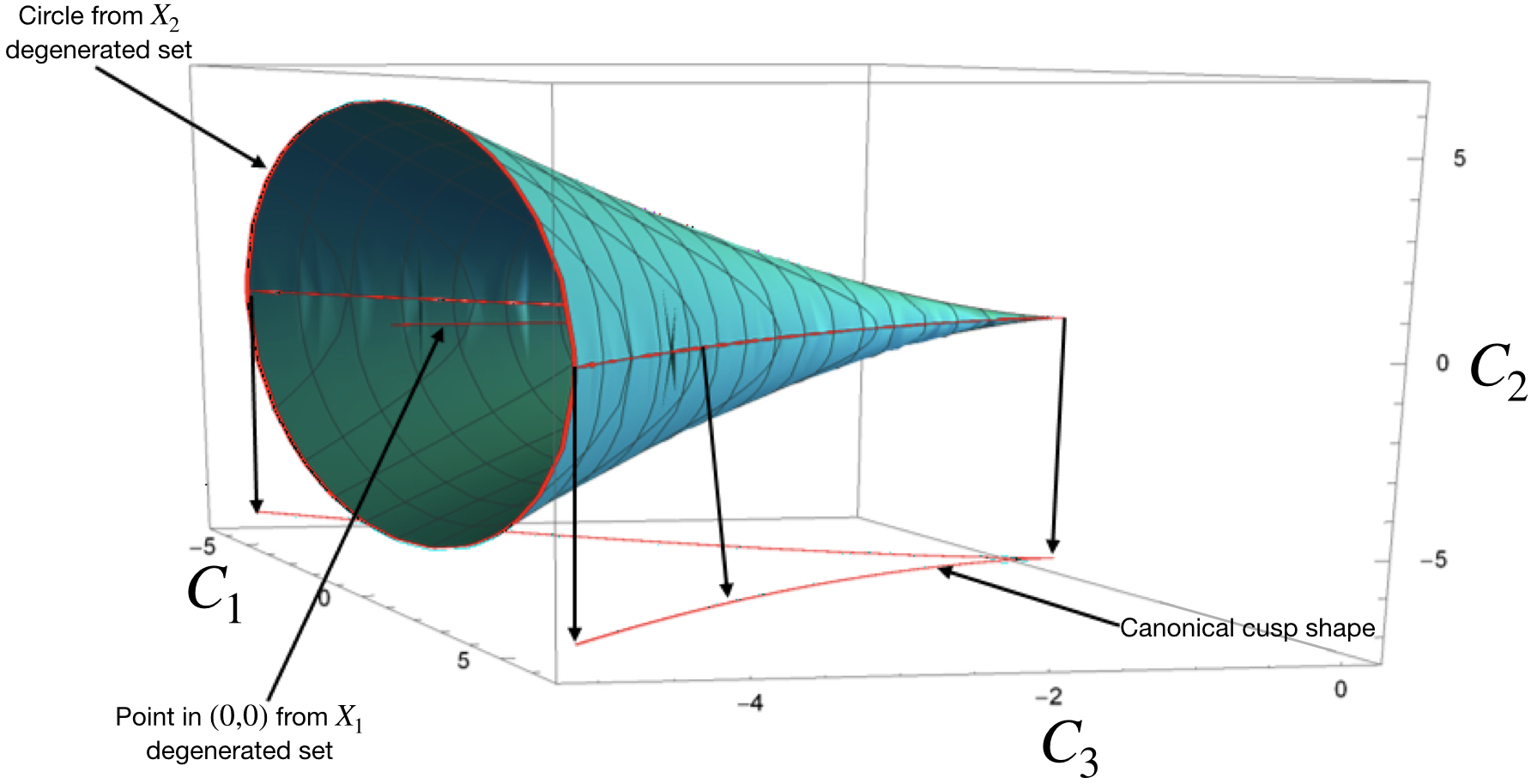}
\caption{\label{fig8} Caustic of a pure rotationally symmetric optical system mapped in the control space. The two solutions $X_1$ and $X_2$ are reproduced after considering only the influence of third order spherical aberration.}
\end{figure}

\noindent The caustic generated by the $X_2$ degenerated set will then be a circle of radius $\sqrt{-8C_3^3/27}$ in the control space. These two simple examples provided aid in understanding the method's immense potential since the analytical technique provides very quickly new informations while obtaining the same optical behaviours that were previously observed using complex numerical computations. Fig.~\ref{fig8} shows the evolution of the caustic observed in a rotationally symmetric optics only affected by the spherical aberration, and considering an interval of the control variable $C_3\in[-5,0]$. Applying J. Nye's method \cite{nye_relation_2005}, one can demonstrate that the radius of the caustic circle in the control space $(C_2,C_3)$ will obey the equation of a canonical cusp defined in Thom's theory (see table \ref{tab:elementary catastrophes}). The apex of this cusp is located at position $(0,0,0)$ of the tridimensional control space. Therefore, following \cite{nye_relation_2005}, we find that the position given by $C_3 = 0$ corresponds to the Gaussian image plane along the optical axis $z_g = -u_{\alpha a}^2/C_{df}$. The cusp profile projected onto the $(C_2,C_3)$ control space is also shown on Fig.~\ref{fig8}. 

\subsubsection{Introduction of tilt along the optical axis: effect on the canonical cusp in real space}

J. Nye also investigated the orientation and the distortion of this cusp in the control space arising from an optical symmetry breaking or a non-dissipative propagation within a general anisotropic and inhomogeneous medium \cite{nye_orientations_1984}. The same procedure can be applied in CPO considering - for instance - the introduction of dipolar field $\Psi_1$ along the optical axis. From \eqref{Astigmatismeintegralequation} we know that such dipolar induces an $A_0$ coefficient which has to be included in the control variables through the expressions \eqref{eq:control variables annexes 1} to \eqref{eq:control variables annexes 9}. We notice that the generating function remains defined by only the three first control variables. Therefore, the caustic equation of the $X_2$ degenerated set must follow the same equation \eqref{causticcircle}. In other words, nothing change within the control space due to the introduction of dipolar field. 

Nevertheless, the effect of $A_0$ can be understood when switching from the control space to the real space. Performing this change while keeping $A_0=0$ first, one simply has to rewrite equation \eqref{causticcircle} by replacing $C_1$ and $C_2$ using their expressions \eqref{eq:control variables annexes 1} and \eqref{eq:control variables annexes 2}:
\begin{equation}
    x^2+y^2 = -\frac{1}{27}\frac{(C_{df}z+u_{\alpha a}^2)^3}{u_{\alpha a}^2C_sz}
\end{equation}

\noindent The same procedure is then performed when adding $A_0$ in $C_1$ and $C_2$. Equation \eqref{causticcircle} then reads: 
\begin{empheq}[left={\empheqlbrace\,}]{align}\label{eq:circle caustics in real plane with A_0}
    \begin{split}
        (x-\Re(\zeta))^2&+(y-\Im(\zeta))^2 \\
        &= -\frac{1}{27}\frac{(C_{df}z+u_{\alpha a}^2)^3}{u_{\alpha a}^2C_sz}
    \end{split}\\
        \zeta &= \frac{A_0z}{u_{\alpha a}}e^{i\chi_a}
\end{empheq}

\noindent To plot these equations and qualitatively observe the impact of $A_0$ on the caustic, we have arbitrarily set $-C_{df} = C_s = u_{\alpha a} = 1$. The results are reported on Fig.~\ref{fig8}(a) for $A_0=0$, and Fig.~\ref{fig8}(b) for an arbitrary value $A_0=-1$. These results are in agreement with the well-known observation that a dipole field $\Psi_1$ at first order simply tilts the optical axis. The overall geometry of the caustic remains unchanged, as it was already obvious from the control space perspective.  

\begin{figure}
\centering
\includegraphics[width=\columnwidth]{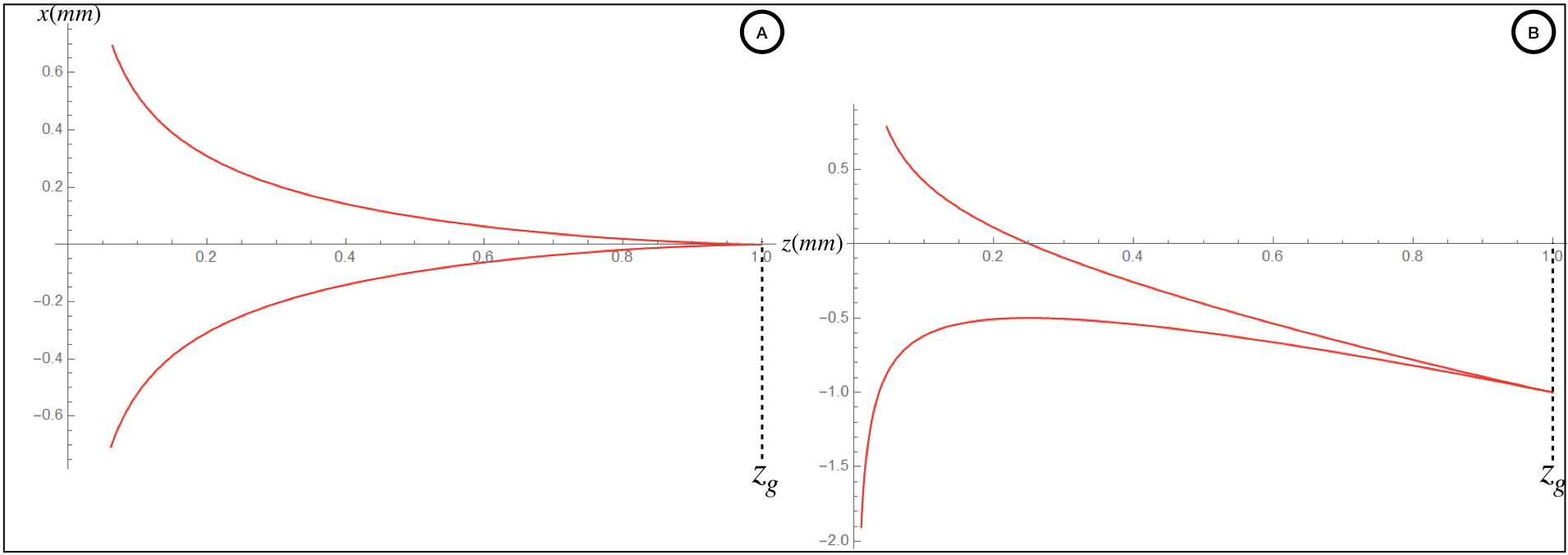}
\caption{\label{fig9} Real space profile of the cusp catastrophe observed in a rotationally symmetric system with the presence of spherical aberration. A : with and B : without deflection introduced through the presence of a dipolar field component $\Psi_1$.}
\end{figure}

\subsubsection{System combining fields with rotational and quadrupole symmetries}

As previously done, one can add a quadrupole field $\Psi_2$ which will induce a two-fold astigmatism term $A_1$ linked to $\Psi_2$ through \eqref{Astigmatismeintegralequation}. However, unlike the previous analysis that used optical calculations of $A_1$ knowing the value of $\Psi_2$, we now simply use the catastrophe theory formalism to investigate the influence of this aberration, without any \textit{a priori} knowledge of the real fields. The methodology will then be identical to that used in the previous section for the rotationally symmetric system and will consist of analysing the geometrical behaviour of the generating function within the control space. 

It is crucial to note that the addition of $A_1$ results in an increased number of control variables (see equations \eqref{eq:control variables annexes 1} to \eqref{eq:control variables annexes 9}), making the equations denser and more challenging to interpret at a first sight. We will see for instance that the determinant of the Hessian will be written as a fourth order polynomial function. Finding the roots of this polynomial is still mathematically possible, but much more complex than in the previous quadratic situation. We have nevertheless chosen to put all the demonstrations in the main body of the paper, because here lies the beauty of this analytical approach. This makes it more difficult to read, but we shall see that - in practice - this is only due to cumbersome expressions The essence of the reasoning remains however as simple as in the previous situation. Practically, the analytical expressions for the roots have been determined with the help of \texttt{Mathematica} software. The code is also given in the supplementary material for information.\\

Thus, we now examine an optical system with only $C_s$ and $A_1$ aberration coefficients. The generating function can be written : 
\begin{equation}\label{eq:generating function C_4 and C_5}
    \begin{split}
        \Phi(x'_a,y'_a,C'_1,C'_2,&C_3,C'_4,C'_5) = C'_1x'_a+C'_2y'_a\\
        &+C_3(x_a^{'2}+y_a^{'2})+C'_4(x_a^{'2}-y_a^{'2})\\
        &+C'_5x'_ay'_a+x_a^{'4}+2x_a^{'2}y_a^{'2}+y_a^{'4}
    \end{split}
\end{equation}

\noindent where we have noted state and control variables with a prime corresponding to an arbitrary coordinate system relative to the caustic orientation in order to simplify subsequent expressions. Only the $C_3$ coefficient remains without prime as it is not related to the caustic orientation and will then remains unchanged in the following simplifications. Using usual polar coordinates $r= \sqrt{x_a^{'2}+y_a^{'2}}$ and $\theta$ corresponding to the argument, the generating function reads:
\begin{equation} \label{generatingpolarcoordinate}
    \begin{split}
        \Phi(r,\theta,C'_1,C'_2,&C_3,C'_4,C'_5) = C'_1r\text{cos}(\theta)+C'_2r\text{sin}(\theta)\\
        &+C_3r^{2}+C'_4(r^{2}\text{cos}^2(\theta)-r^{2}\text{sin}^2(\theta))\\
        &+C'_5r^{2}\text{cos}(\theta)\text{sin}(\theta)+r^{4}
    \end{split}
\end{equation}

\begin{figure}
\centering
\includegraphics[width=\columnwidth]{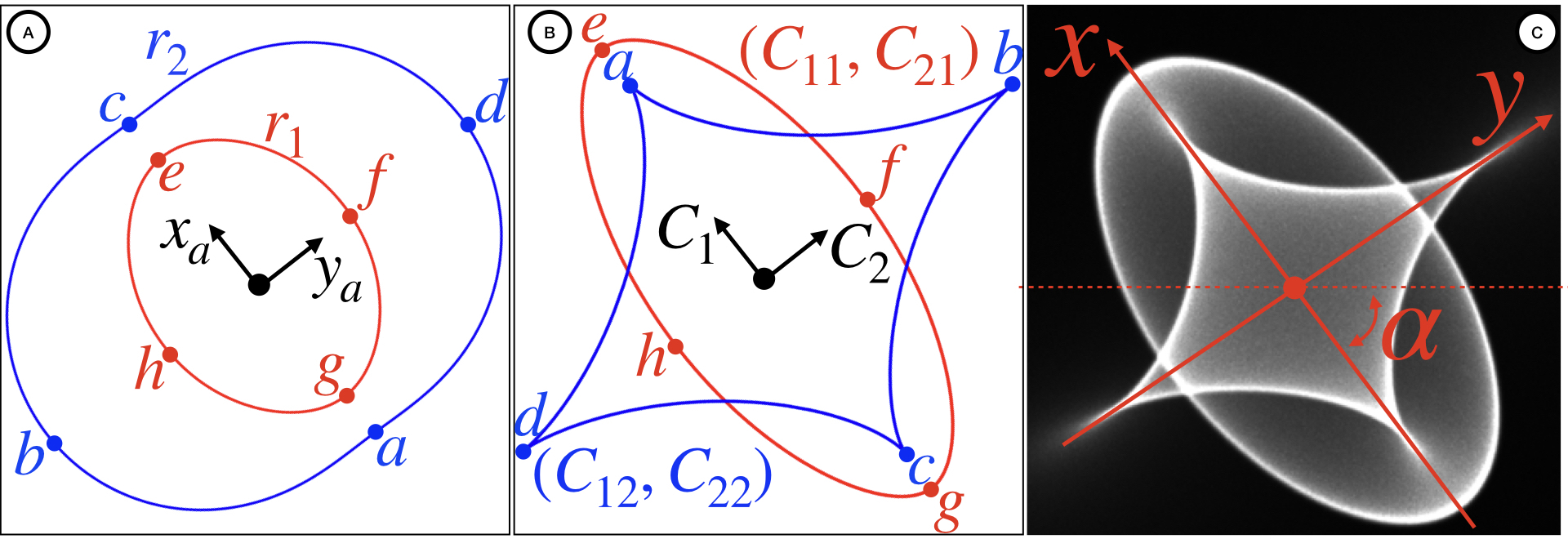}
\caption{\label{fig10} A: Map of the Hessian determinant degenerated sets in the space defined by the two state variables $x_a$ and $y_a$. B: Caustics obtained in the control space related to the two degenerated solutions (red and blue) depicted in A. C : Experimental caustic observed in the I2TEM microscope.
}
\end{figure}

\noindent By linking a set of new control variables to a basis defined by the caustic's symmetry, one can in practice lower the number of effective variables and simplify the problem. Indeed, Fig.~\ref{fig10}(c) reports an example of caustic observed in the I2TEM when a quadrupole field is applied with the condensor stigmators. As shown, the two new coordinates axes $(x,y)$ are tilted relative to the global image axis $(x',y')$ (again the prime notations are used when orientation is unrelated to the caustic) following the natural orientation of the caustic. We have then to define the set of state and control variables attached to these main caustic axis - keeping in mind that the caustic could be rotated when $A_1$ changes. This is the origin of the two control variables $C'_4$ and $C'_5$ which physically represent this same influence emerging from the two-fold astigmatism. $A_1$ being a complex number, equations \eqref{eq:control variables annexes 4} and \eqref{eq:control variables annexes 5} shows that one needs two control variables to define the action of $A_1$ in an arbitrary reference system through its real and imaginary parts. The modulus of the complex interplay of $C'_4$ and $C'_5$ can then be represented by a new control variable $C_4$. This new control variable is associated to the caustic coordinate system independently of the value of $\alpha$ - which then correspond to the argument of this complex association. Hence, from \eqref{eq:control variables annexes 4} and \eqref{eq:control variables annexes 5} one obtains:
\begin{eqnarray}
   C_4 &=& -\sqrt{C_4^{'2}+\left( \frac{C'_5}{2}\right)^2} = -\frac{2|A_1|}{C_s}u_{\alpha a}^2 \label{newc4variable}\\
        \alpha &=& -\frac{\text{arctan}\left(\tfrac{C'_5}{2C'_4}\right)}{2}=-\frac{\text{arg}(A_1)+2\chi_a}{2}
\end{eqnarray}

\noindent With those definitions, one can then simplify the generating function \eqref{generatingpolarcoordinate} after applying the following simplification:
\begin{equation}
\begin{split}
  C'_4(r^{2}\text{cos}^2(\theta)&-r^{2}\text{sin}^2(\theta))\\
  &+C'_5r^{2}\text{cos}(\theta)\text{sin}(\theta) = C_4(x_a^2-y_a^2)   
\end{split}
\end{equation}

\noindent with:
\begin{empheq}[left={\empheqlbrace\,}]{align}
        x_a &= r\text{cos}(\theta+\alpha)\\
        y_a &= r\text{sin}(\theta+\alpha)
\end{empheq}

\noindent Finally, we define the two new control variables $(C_1,C_2)$ within the new coordinate system tilted by $\alpha$ (see Fig.~\ref{fig10}(b)):
\begin{empheq}[left={\empheqlbrace\,}]{align}
        C_1 &= C_1'\text{cos}(\alpha)+C_2'\text{sin}(\alpha)\\
        C_2 &= C_2'\text{cos}(\alpha)-C1'\text{sin}(\alpha)
\end{empheq}

\noindent with these new coordinates, one can reconstruct the generating function in a more condensed manner using only four control variables, see Fig.~\ref{fig10}(a) and Fig.~\ref{fig9}(b):
\begin{equation}
\label{generatingastigmatism}
   \begin{split}
        \Phi(x_a,y_a,&C_1,C_2,C_3,C_4) = C_1x_a+C_2y_a+C_3(x_a^2+y_a^2)\\
        &+C_4(x_a^2-y_a^2)+x_a^4+2x_a^2y_a^2+y_a^4
     \end{split}
\end{equation} 

\noindent Additionally, one can perform all the analysis within the restricted range $C_4<0$. Indeed, considering positive values of $C_4$ will be mathematically identical, since the generating function remains the same by simply swapping the state and control space axis i.e. $x_a\rightarrow y_a$, $y_a\rightarrow x_a$,$C_1\rightarrow C_2$, $C_2\rightarrow C_1$ and $C_4\rightarrow -C_4$. The critical set can now be extracted:

\begin{empheq}[left={\empheqlbrace\,}]{align}
        \frac{\partial \Phi}{\partial x_a} = C_1+2x_a(C_3+C_4)+4x_a^3+2x_ay_a^2 = 0\\
        \frac{\partial \Phi}{\partial y_a} = C_2+2y_a(C_3-C_4)+2x_a^2y_a+4y_a^3 = 0
\end{empheq}

\noindent and the degenerated set:
\begin{equation}
    \frac{\partial f}{\partial x_a}\frac{\partial \Phi}{\partial y_a}-\left(\frac{\partial \Phi}{\partial x_a\partial y_a}\right)^2 = 0
\end{equation}

\noindent are found by extracting the roots of this fourth order polynomial function:
\begin{equation}
    \begin{split}
        48x_a^4+96x_a^2y_a^2+&48y_a^4+32C_3(x_a^2+y_a^2)\\
        &-16C_4(x_a^2-y_a^2)+4(C_3^2-C_4^2) = 0
    \end{split}
\end{equation}

\noindent which is more conveniently written in polar coordinates:

\begin{equation}
    48r^4+(32C_3-16C_4\text{cos}(2\theta))r^2+4(C_3^2-C_4^2) = 0
\end{equation}

\noindent Considering $X = r^2$, the determinant of the Hessian corresponds to a second order polynomial function, which roots can easily be found. Using our \texttt{Mathematica} script given in supplementary material, we obtain the two degenerated sets: 

\begin{widetext}
\begin{eqnarray}\label{eq:solutions r_i A_1 caustics}
        r_1 &=& \frac{\sqrt{-2C_3+C_4\text{cos}(2\theta)-\sqrt{C_3^2+3C_4^2-4C_3C_4\text{cos}(2\theta)+C_4^2\text{cos}^2(2\theta)}}}{\sqrt{6}}\\
        r_2 &=& \frac{\sqrt{-2C_3+C_4\text{cos}(2\theta)+\sqrt{C_3^2+3C_4^2-4C_3C_4\text{cos}(2\theta)+C_4^2\text{cos}^2(2\theta)}}}{\sqrt{6}}
\end{eqnarray}
\end{widetext}

\noindent These solutions are shown on Fig.~\ref{fig10}(a), where the red (respectively blue) curve corresponds to $r_1$ ($r_2$). From these solutions one can extract the caustic shape within the control space by solving the Fermat's principle. As done previously, we define a parametric equation for each solutions expressed in ($C_{1i}$, $C_{2i}$) for the set $r_i$:
\begin{widetext}
\begin{align}
\begin{split}
         C_{11} &= \frac{\sqrt{-4C_3+2C_4\text{cos}(2\theta)-\sqrt{2}\sqrt{2C_3^2+7C_4^2-8C_3C_4\text{cos}(2\theta)+C_4^2\text{cos}(4\theta)}}}{6\sqrt{3}}\\
         &\times\left(-2C_3-6C_4-2C_4\text{cos}(2\theta)+\sqrt{2}\sqrt{2C_3^2+7C_4^2-8C_3C_4\text{cos}(2\theta)+C_4^2\text{cos}(4\theta)} \right)\text{cos}(\theta)
\end{split}\label{eq:C_11 A_1 caustics}\\         
\begin{split}
         C_{21} &= \frac{\sqrt{-4C_3+2C_4\text{cos}(2\theta)-\sqrt{2}\sqrt{2C_3^2+7C_4^2-8C_3C_4\text{cos}(2\theta)+C_4^2\text{cos}(4\theta)}}}{6\sqrt{3}}\\ 
         &\times\left(-2C_3+6C_4-2C_4\text{cos}(2\theta)+\sqrt{2}\sqrt{2C_3^2+7C_4^2-8C_3C_4\text{cos}(2\theta)+C_4^2\text{cos}(4\theta)} \right)\text{sin}(\theta)
\end{split} \label{eq:C_21 A_1 caustics} \\
\begin{split}
        C_{12} &= -\frac{\sqrt{-4C_3+2C_4\text{cos}(2\theta)+\sqrt{2}\sqrt{2C_3^2+7C_4^2-8C_3C_4\text{cos}(2\theta)+C_4^2\text{cos}(4\theta)}}}{6\sqrt{3}}\\
        &\times\left(2C_3+6C_4+2C_4\text{cos}(2\theta)+\sqrt{2}\sqrt{2C_3^2+7C_4^2-8C_3C_4\text{cos}(2\theta)+C_4^2\text{cos}(4\theta)} \right)\text{cos}(\theta)
\end{split} \label{eq:C_12 A_1 caustics}\\
\begin{split}
        C_{22} &= -\frac{\sqrt{-4C_3+2C_4\text{cos}(2\theta)+\sqrt{2}\sqrt{2C_3^2+7C_4^2-8C_3C_4\text{cos}(2\theta)+C_4^2\text{cos}(4\theta)}}}{6\sqrt{3}}\\ 
        &\times\left(2C_3-6C_4+2C_4\text{cos}(2\theta)+\sqrt{2}\sqrt{2C_3^2+7C_4^2-8C_3C_4\text{cos}(2\theta)+C_4^2\text{cos}(4\theta)} \right)\text{sin}(\theta)
\end{split} \label{eq:C_22 A_1 caustics} 
\end{align}
\end{widetext}

\noindent Two pairs of parametric equations are then provided by the catastrophe theory, enabling us to examine an optical system containing quadrupolar and rotationally symmetric fields without ever requiring a quantitative estimate of their values. Yet, we are aware that this is still feasible in the present case only because the polynomial is below order five - which enables analytical solutions for the roots. Nevertheless, the situation studied here is, in practice, what is mostly observed in a electron and ion microscopy experiments. The main exceptions being the cases where an aberration corrector drastically reduces the contribution of $C_s$ so that the introduction of aberrations $B_2$, $A_2$, and higher becomes necessary.\\

We will now look at the information that can be extracted from this pair of solutions. We will see that some interesting details that would have been difficult, if not impossible, to find using the numerical method, can be extracted from these solutions. The caustic generated using $r_1$ falls within the classic definition of a \emph{fold} while the one generated using $r_2$ set creates four \emph{cusps} in the control space (see Fig.~\ref{fig10}) \cite{zeeman_catastrophe_1977}. To facilitate the discussion, in the following we will refer to them as \emph{fold-caustic} and \emph{cusp-caustic} respectively.

The correspondence between the rays positions within the two caustics and their locations in the pupil plane can now be established. To do so, let's define $\phi$, the angle associated to a random point ($C_{1i},C_{2i}$) located on one of the two caustics solutions $i=(1,2)$ and relatively to the control space main axis $(C_1,C_2)$. Equations  \eqref{eq:C_11 A_1 caustics}, \eqref{eq:C_21 A_1 caustics}, \eqref{eq:C_12 A_1 caustics} and \eqref{eq:C_22 A_1 caustics} can then be simply written as $C_{1i} = g_{(1i)}\text{cos}(\theta),C_{2i} = g_{(2i)}\text{sin}(\theta)$, where $g_{(ni)}$ denote monotonic functions. We then obtain the following relations between some specific orientations of the degenerated set in the pupil plane and the two caustics in the control space: 
\begin{equation}
    \begin{gathered}
        \theta\in\left\{-\frac{\pi}{2},\frac{\pi}{2} \right\}\implies C_{1i} = 0\implies\psi\in\left\{-\frac{\pi}{2},\frac{\pi}{2} \right\}\\
        \theta\in\left\{0,\pi\right\}\implies C_{2i} = 0\implies\psi\in\left\{0,\pi\right\}
    \end{gathered}
\end{equation}

\noindent The exact locations of the cardinal points - which corresponds to the points along the two axis of symmetry - depend on the sign of the control variables $C_{ji}(\theta)$. We can show that in most cases we have $\theta = \psi$. These cardinal points are highlighted on Fig.~\ref{fig9}(a) and Fig.~\ref{fig9}(b) by letters ranging from $b,d,e,f,g$. An inversion is observed for the cusp-caustic when $\theta = \pm\pi/2$ leading to $\psi= \mp\pi/2$. These two specific locations are noted $a,c$ on Fig.~\ref{fig9}(a) and Fig.~\ref{fig9}(b). Additionally, one can show that these cardinal points can be specified only for a certain range of $C_3$. In particular, these cardinal points can be defined for the fold-caustic only if $C_3<C_4$ but only when $C_3<-C_4$ for the cusp-caustic. Going back to the real space, these two limits simply represent the sagittal and the tangential focal points defined $z_{s/t} = -\frac{u_{\alpha a}^2}{\pm|A_1|+C_{df}}$.\\ 

\begin{figure}
\centering
\includegraphics[width=0.5\textwidth]{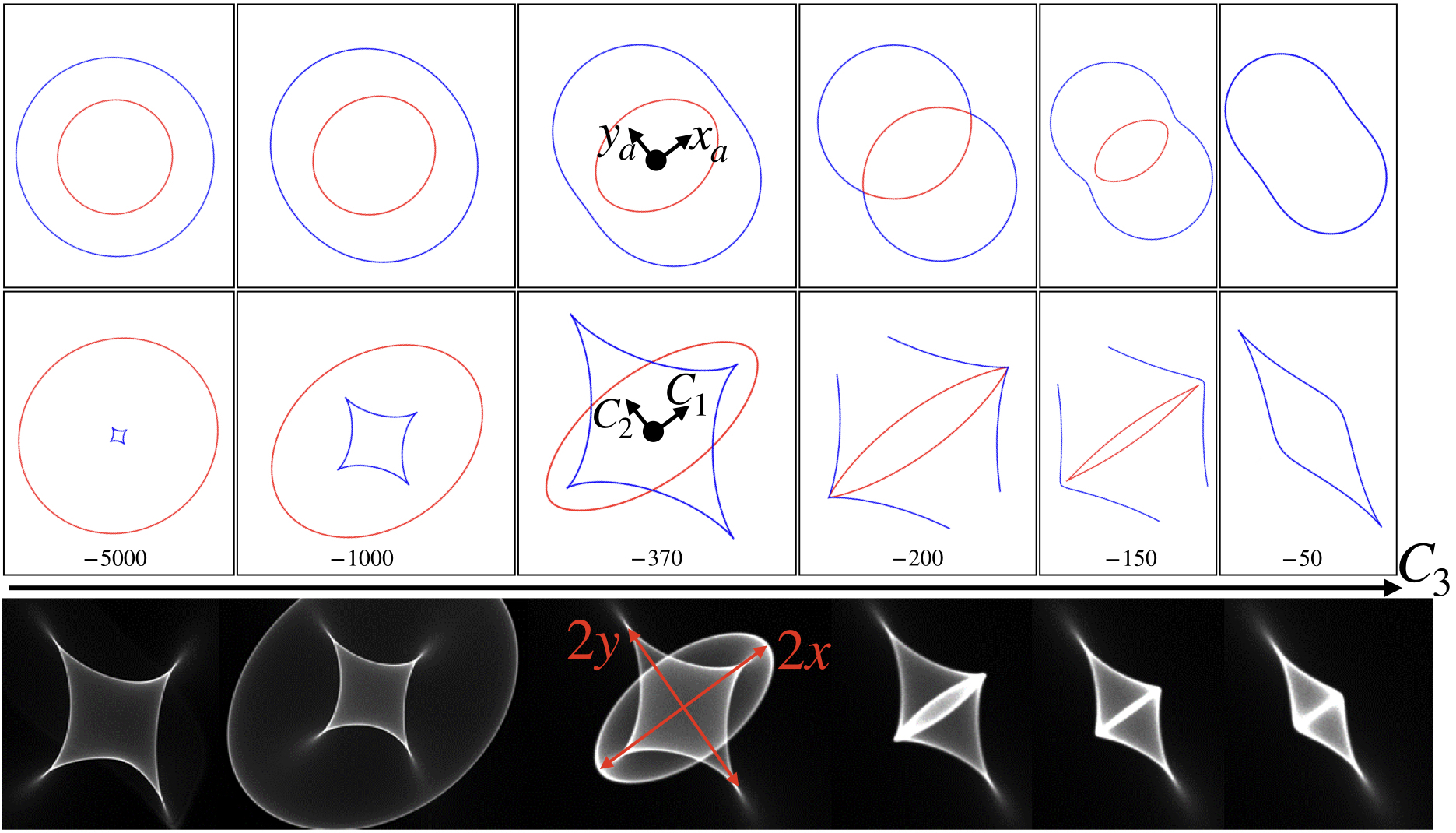}
\caption{\label{fig11} A: Evolution of the Hessian determinant degenerated sets with $C_3$ control variable. B: Evolution of the associated caustics in the control space. C : Evolution of the experimental caustic observed in the I2TEM microscope by changing the defocus parameter $C_{df}$, linked to the $C_3$ control variable through (\ref{C3control}).}
\end{figure}

The general behaviour of these caustics as a function of $C_3$ can be intuitively understood by plotting them - which constitutes a quick and easy task from the previously-derived analytical solutions. For example, Fig.~\ref{fig10} displays the evolution of a caustic within the $(C_1,C_2)$ control space calculated for an arbitrary two-fold astigmatism represented by the control variable $C_4=-100$. The caustic is observed for different values of $C_3$, which physically corresponds to a translation along the optical axis, i.e. a defocus. Experimental data are also shown for a qualitative comparison. We have chosen not to include any scales, as these would be of no real physical interest. Indeed, as it was discussed earlier, in this catastrophe theory method, we are only concerned with the evolution of the relative dimensions and symmetries between the morphologies of the two solutions within the control space. Keeping this in mind, we can assess the agreement between the plots of these solutions in $(C_1,C_2)$ space as a function of $C_3$ and the experimental data acquired with the I2TEM microscope - for a non-zero astigmatism $A_1$. The caustics evolution has been acquired by slightly changing the focal strength of the condenser system (in the present example, the first condenser lens).\\ 

By developing the two solutions using the $(C_{1i},C_{2i})$ parametric equations \eqref{eq:C_11 A_1 caustics} to \eqref{eq:C_22 A_1 caustics}, one can obtain exact relationships linking measurable geometrical parameters of the caustic - extracted for specific $C_3$ parameters. Such relationships, which depend on the aberration coefficients, can be fairly simple. The scenario defined by $C_3 = 2C_4$, for example, is relevant to some practical considerations. Indeed, along the $C_1$ axis, the fold-caustic and the cusp-caustic eventually share two cardinal points at this specific $C_3 = 2C_4$ location. The scenario is clearly observed on Fig.~\ref{fig10} for $C_3=-200$. At this specific point we can extract the relationship $C_{11}(\theta = 0) = C_{12}(\theta = 0) = -2C_4\sqrt{-2C_4}$, which can be written in real space as:
\begin{equation}\label{c32c4cond}
    \frac{x}{z} = \frac{4|A_1|^{\frac{3}{2}}}{|u_{\alpha a}|\sqrt{C_s}}
\end{equation} 

\noindent Besides, the condition $C_3 = 2C_4$ between the control parameters implies that $-2z(C_{df}+2|A_1|) = 1$. Thus, we could for instance design an experiment where a two-fold astigmatism $A_1$ and a defocus $C_{df}$ are manually introduced to match this condition $C_3 = 2C_4$ - the proper value for the defocus corresponding to a contact between the "cusp-caustic" and the "fold-caustic". Then, one can easily find the value of this defocus, which gives access to the value of $A_1$. Knowing this value, and using the relationship \eqref{c32c4cond}, one can precisely extract the spherical aberration parameter $C_s$. To do so however, one needs to eliminate the term $u_{\alpha a}$ - corresponding to the position of the marginal ray in the pupil plane - through e.g. two subsequent measurements using two different condenser apertures. 

We have thus demonstrated that a simple relation between the coefficients could be derived using catastrophe theory, which would have been challenging to extract using a brute force numerical approach. The condition $C_3 = 2C_4$ is however quite specific. We will now work out another simple relation in a more general situation. Let's examine the set of positions in the control space - along the optical axis - defined by $C_3<2C_4$, as shown on  Fig.~\ref{fig10}. The relations \eqref{eq:C_11 A_1 caustics} to \eqref{eq:C_22 A_1 caustics} enable us to establish a relationship between the two perpendicular directions $x,y$ given by the larger dimensions of the fold and cusp caustics, respectively:
\begin{equation}
\label{c3mc2c4relation}
    \frac{C_{11}(\theta = 0)}{C_{22}(\theta = -\pi/2)} = \frac{C_3+C_4}{2\sqrt{3}C_4}
\end{equation}

\noindent Therefore, a simple measurement of these distances leads to the following relation between the defocus and two-fold astigmatism:
\begin{equation}\label{CdfA1xyrelation}
    \frac{x}{y} = \frac{2z(C_{df}-|A_1|)+2u_{\alpha a}}{-6\sqrt{2}|A_1|}
\end{equation}

In summary, we have chosen to highlight two specific situations - encapsulated in equations \eqref{c32c4cond} and \eqref{CdfA1xyrelation} - to illustrate the strength of CCPO. Although we haven't used these relationships experimentally yet, we intend to conduct specific TEM experiments to test our approach before implementing it on a FIB instrument - where measuring aberrations is a significant technological difficulty.

The aim of this work was to show the great potential offered by the mathematical framework of catastrophe theory. First of all, and this is far more important to us than any applications, the approach developed by R. Thom opens up a completely new window of opportunity for understanding these complex figures that we all observe in our every day. Furthermore, the universal character of this theory enable us to make very meaningful analogies between the field of optics and other fields of physical science where bifurcations can also be observed.

Indeed, the relations obtain in the latter part of this article - namely \eqref{c32c4cond} and \eqref{CdfA1xyrelation} - convey a very universal character as they must also be valid in any system characterized by the same generating function. In the context of CPO, they offer a fresh and new insight into the detection of aberrations, for instance in focused ion beam systems.

It must be stressed that such a purely analytical approach is however limited to a small number of aberration coefficients - since from order five onwards it would no longer be possible to define the roots of the determinant of the Hessian univocally. Nevertheless, we believe that by playing with experimentally free parameters - such as defocus $C_{df}$, two-fold $A_1$ and three-fold $A_2$ astigmatisms - it should be always possible to define a situation where the dimensionality of the control space remains restricted below five. 

\section{Conclusion}

We applied the formalism of catastrophe theory to a charged particle optics problem by adapting the classical eikonal theory into the geometrical language originally developed by R. Thom. This formalism allowed us to reproduce the behaviour of optical caustics analytically without any \textit{a priori} knowledge of our optics. We were able to gauge the efficiency of this innovative method by comparing these results with those obtained directly from the eikonal formalism. We were thus able to reproduce the caustics observed in a TEM containing spherical aberration and axial astigmatism without having to calculate the distribution of electromagnetic potentials inside each optical element. It is in this highly computational phase that traditional methods require strong resources. Indeed, estimating these potentials remains unfortunately necessary if caustics have to be evaluated by the eikonal method and a fortiori using conventional ray tracing. For example, these calculations are too time-consuming to be used for actual measurements during investigations if one wishes to rapidly extract the aberrations from caustics in order to optimize the alignment of an instrument. The new mathematical expressions established during this work offer a whole new paradigm because their estimations are almost instantaneous. The geometric approach of catastrophe theory should then enable us in the future to perform quick measurements of the optical properties of various systems using charged particles. We are now mainly considering applications to electrostatic optics which are the heart of ion microscopes like FIB or SIMS. The nature of ions makes the application of usual optical characterization methods, widely used in electron optics, very complex, if not impossible. Caustics observed by etching and then analysed by catastrophe theory take a few milliseconds, paving the way for live measurement of the optical properties in future ions-based instruments, which is a first step towards optimizing their optics.

\begin{acknowledgments}
\noindent This study has been supported through the grant NanoX n° ANR-17-EURE-0009 in the framework of the "Programme des Investissements d’Avenir". We also acknowledge financial support by the European Union's Horizon Europe framework program
for research and innovation under grant agreement n° 101094299 (IMPRESS project).
\end{acknowledgments}

\bibliography{cpo.bib}
\bibliographystyle{unsrt}
\end{document}